\documentclass[12pt]{article}
\pdfoutput=1
\usepackage{epsf,amsfonts,amssymb,epsfig,amsmath,mathtools,graphics,slashed}
\usepackage{hyperref,hep,graphicx,subfig}

\makeatletter

\newcommand{\Rmnum}[1]{\expandafter\@slowromancap\romannumeral #1@}
\makeatother

\newcommand{\nn}{\notag \\}

\begin{document}

\makeatletter
\renewcommand{\theequation}{\thesection.\arabic{equation}}
\@addtoreset{equation}{section}
\makeatother

\baselineskip 18pt

\begin{titlepage}

\vfill

\begin{flushright}
Imperial/TP/2013/JG/03\\
\end{flushright}

\vfill

\begin{center}
   \baselineskip=16pt
   {\Large\bf Competing $p$-wave orders}
  \vskip 1.5cm
  \vskip 1.5cm
      Aristomenis Donos$^1$, Jerome P. Gauntlett$^2$ and Christiana Pantelidou$^2$\\
   \vskip .6cm
    \vskip .6cm
      \begin{small}
      \textit{$^1$DAMTP, 
       University of Cambridge\\ Cambridge, CB3 0WA, U.K.}
        \end{small}\\
      \vskip .6cm
      \begin{small}
      \textit{$^2$Blackett Laboratory, 
        Imperial College\\ London, SW7 2AZ, U.K.}
        \end{small}\\*[.6cm]

\end{center}

\vfill

\begin{center}
\textbf{Abstract}
\end{center}

\begin{quote}
We construct electrically charged, asymptotically $AdS_5$ black hole solutions
that are dual to $d=4$ CFTs in a superfluid phase with either $p$-wave or $(p+ip)$-wave order.  
The two types of black holes have non-vanishing charged two-form in the bulk and appear at the same
critical temperature in the unbroken phase. Both the $p$-wave and the $(p+ip)$-wave phase can be thermodynamically preferred,
depending on the mass and charge of the two-form, and there can also be first order transitions
between them. The $p$-wave black holes have a helical structure and some
of them exhibit the phenomenon of pitch inversion as the temperature is decreased.
Both the $p$-wave and the $(p+ip)$-wave black holes have zero entropy density ground states at zero temperature
and we identify some new ground states which exhibit scaling symmetry, including a 
novel scenario for the emergence of conformal symmetry in the IR.
\end{quote}

\vfill

\end{titlepage}
\setcounter{equation}{0}


\section{Introduction}
Investigating the thermal properties of strongly coupled matter using holographic techniques 
has revealed an extraordinarily rich landscape of novel black hole solutions and it seems likely 
that we have still only seen the tip of the iceberg.  

An interesting class of examples involve strongly coupled conformal field theories (CFTs) in flat spacetime when held at finite chemical potential with respect to a global $U(1)$ symmetry. These are described by electrically charged, asymptotically AdS black holes with event horizons that are topologically planar.
Within this class one finds black holes that are dual to superfluid phases which spontaneously break the $U(1)$ symmetry, with $s$-wave \cite{Gubser:2008px,Hartnoll:2008vx,Hartnoll:2008kx}, $p$-wave \cite{Gubser:2008zu,Gubser:2008wv,Roberts:2008ns,Aprile:2010ge,Ammon:2009xh} and
$d$-wave \cite{Chen:2010mk,Benini:2010pr,Kim:2013oba} order. In addition there are phases that spontaneously break some of the
Poincar\'e symmetries of the boundary theory, including helical current phases \cite{Nakamura:2009tf,Donos:2012wi} and striped phases \cite{Donos:2011bh,Rozali:2012es,Donos:2013wia,Withers:2013loa,Withers:2013kva}. These latter solutions break parity and time reversal invariance, but the holographic 
charge density waves of \cite{Donos:2013gda} do not. Furthermore, it is possible to simultaneously break both the $U(1)$ symmetry 
and the Poincar\'e symmetry as in the helical $p$-wave and $(p+ip)$-wave superconducting phases \cite{Donos:2011ff,Donos:2012gg}.
A further investigation of this latter class of examples will be the focus of this paper.

As in \cite{Donos:2011ff,Donos:2012gg} we will study a model in $D=5$ space-time dimensions which couples the metric to an abelian gauge-field
and a charged two-form. The two-form, $C$, has mass $|m|$, charge $e$ and 
is dual to a self-dual tensor operator in a dual $d=4$
CFT with scaling dimension $\Delta=2+|m|$. The high temperature phase of the CFT with non-vanishing chemical
potential is spatially homogeneous and isotropic and is 
described by the electrically charged AdS-RN planar black hole solution. 
It was shown in \cite{Donos:2011ff} that the AdS-RN black hole becomes unstable if $e^2>m^2/2$. 
More precisely, it was shown that there are linearised perturbations 
of the charged two-form labelled by wave-number $k$, associated with both $p$-wave and $(p+ip)$-wave
order, which become tachyonic below some critical temperature. The instabilities first appear for 
$k\ne 0$ and moreover, since the linearised perturbations 
are governed by exactly the same ODE, the instabilities for the two types of order appear at the same
critical temperature. 

Some fully back-reacted $p$-wave black hole solutions for this model were constructed in \cite{Donos:2012gg} 
and here we 
will review and extend that work. We will also construct the first back-reacted $(p+ip)$-wave black holes for this model.
It is natural to ask which of the two phases is thermodynamically preferred and how they compete\footnote{Some other holographic studies of competing orders appear in \cite{Basu:2010fa,Donos:2011ut,Donos:2012yu,Musso:2013ija,Cai:2013wma,Nitti:2013xaa,Liu:2013yaa,Nie:2013sda,Amado:2013lia}.}.
For a different class of models, involving $SU(2)$ gauge-fields, and for vanishing wave number, it was shown that the $p$-wave
order is preferred over the $(p+ip)$-wave order \cite{Gubser:2008wv}. By contrast here we will see that depending on the parameters $e,m$ both the helical $p$-wave and the $(p+ip)$-wave superfluid phases can be preferred and there can be first order transitions between them.

The $p$-wave black hole solutions of \cite{Donos:2012gg}, and those that we construct here,
are static and spatially homogenous with a helical, Bianchi $VII_0$ symmetry.
They appear in two-parameter families specified by the temperature, $T$, and
the wave-number, $k$, which fixes the pitch (i.e. the periodicity) 
of the helical order to be $2\pi/k$. At fixed $T$ the thermodynamically preferred
black holes are determined by minimising the free energy density with respect to $k$. 
The numerical results in \cite{Donos:2012gg}
indicated that this condition is equivalent to a simple constraint on the boundary data. Here we will directly prove this result, which also follows from the
general results recently obtained in \cite{Donos:2013cka} concerning the thermodynamics of periodic black brane solutions.
As in \cite{Donos:2012gg}, we find that the pitch of the helix monotonically increases as the temperature is initially lowered away from
the critical temperature. For certain values of $e,m$ we find $p$-wave black holes which exhibit the phenomenon of pitch inversion:
$k$ starts off positive, decreases down to zero, changes sign and then increases in magnitude as the temperature is lowered. 
Such a phenomenon is seen\footnote{It is worth pointing out that such a phenomenon is not present in the
black holes dual to helical current phases that were studied in \cite{Nakamura:2009tf,Donos:2012wi}.}, 
for example, in some chiral nematic liquid crystals and helimagnets. 
In the zero temperature limit, $T\to 0$, the $p$-wave black holes approach zero entropy ground states.
For $k>0$ we find an IR Bianchi VII$_0$ scaling solution as already seen in \cite{Donos:2012gg} (of a very similar type to \cite{Iizuka:2012iv}). 
We also find a new type of zero entropy ground state with $k=0$ and an IR scaling symmetry, somewhat similar to some solutions constructed in \cite{Taylor:2008tg}.
Finally, when $k<0$ we find evidence that
at extremely low temperatures the black holes approach $AdS_5$ in the IR, perturbed by marginal and {\it relevant} $k$-dependent operators. This 
novel emergence of conformal symmetry in the IR is somewhat reminiscent of the recent work on periodic potentials in \cite{Chesler:2013qla}. 

The $(p+ip)$-wave black hole solutions are also labelled by temperature $T$ and a wave-number $k$ associated with a direction we
will label $x_1$. The solutions 
are stationary and preserve two translations, in the $x_2,x_3$ directions, 
as well as translations in the $x_1$ direction when combined with a local gauge transformation.
In addition they preserve one rotation in the $x_2,x_3$ plane. Unlike the $p$-wave black holes, the continuous symmetry for the $(p+ip)$-wave black holes
is the same for both $k=0$ and $k\ne 0$.
These black holes obey some novel Smarr formulae, consistent with the results of \cite{Donos:2013cka}.
The one-parameter family of thermodynamically preferred black holes obtained by minimising the free-energy density 
with respect to $k$, are again specified by simple constraints on the boundary data as expected from \cite{Donos:2013cka}. 
In contrast to the $p$-wave case the stress tensor for the one-parameter family of $(p+ip)$-wave black holes
is that of a spatially homogeneous and isotropic ideal fluid. In the $T\to 0$ limit the entropy density of the $(p+ip)$-wave black holes
goes to zero; we have not managed to extract from the numerics a simple statement about the ground states.
However, we provide some evidence that for some black holes an emergent conformal symmetry again appears.

The plan of the rest of the paper is as follows. Section 2 introduces the $D=5$ gravity model that we study, as well as the
$p$-wave and the $(p+ip)$-wave instabilities of the AdS-RN black hole. The back-reacted $p$-wave and $(p+ip)$-wave 
black holes are discussed in sections 3 and 4, respectively. In section 5 we compare the thermodynamics of the two competing orders
and we briefly conclude in section 6. There is one appendix in which we provide a derivation of the Smarr formula
for the $(p+ip)$-wave black holes.

\section{The model}
We consider a theory of gravity in $D=5$ space-time dimensions coupled to a gauge-field $A$ and
a complex two-form $C$ with Lagrangian density given by the five-form
\begin{equation}
\label{eq:action}
\mathcal{L}=(R+12)\ast 1-\frac{1}{2}\ast F\wedge F-\frac{1}{2}\ast C\wedge \bar{C}-\frac{i}{2 m} C\wedge \bar{H}\,,
\end{equation}
where the bar denotes complex conjugation. The field strengths are given by
\begin{equation}
F=dA,\qquad  H=dC+i e A\wedge C\,.
\end{equation}
The corresponding equations of motion are given by
\begin{align}\label{fulleom}
R_{\mu\nu}&=-4g_{\mu\nu}+\tfrac{1}{2}\left(F_\mu{}^\rho F_{\nu\rho}-\tfrac{1}{6}g_{\mu\nu}F_{\rho\sigma}F^{\rho\sigma}\right)
+\tfrac{1}{2}\left(C_{(\mu}{}^\rho\bar C_{\nu)\rho}-\tfrac{1}{6}g_{\mu\nu}C_{\rho\sigma}\bar C^{\rho\sigma}\right)\,,\nn
d*F&=-\frac{e}{2m}C\wedge \bar C\,,\nn
H&=-im*C\,.
\end{align}

The equations of motion admit a unique $AdS_5$ solution, with $A=C=0$, which is dual to a class of CFTs labelled by the two-parameters 
$e$ and $m$.
The equations of motion also admit the electrically charged AdS-Reissner-Nordstrom black brane solution given by
\begin{equation}
ds^2=-gdt^2 +g^{-1}{dr^2}+r^2(dx_1^2+dx_2^2+dx_3^2)\,,\qquad A=a dt\,,
\end{equation}
with $C=0$ and
\begin{equation}
g=r^2-\frac{r_+^4}{r^2}+\frac{\mu^2}{3}(\frac{r_+^4}{r^4}-\frac{r_+^2}{r^2})\,,\qquad a=\mu(1-\frac{r_+^2}{r^2})\,.
\end{equation}
The AdS-RN black hole has temperature $T=(6 r_+^2-\mu^2) /6 \pi r_+$ and describes the high temperature, spatially homogeneous and isotropic phase of the dual CFT when held at finite chemical potential $\mu$ with respect to the global abelian symmetry.

It was shown in \cite{Donos:2011ff} that when $e^2>m^2/2$ these black holes are unstable\footnote{This was shown by analysing instabilities
of the $AdS_2\times \mathbb{R}^3$ solution which arises as the near horizon limit of the $T=0$ AdS-RN black hole
solution. In principle there could be additional instabilities which do not manifest themselves
in this way.} below a critical temperature
corresponding to the formation of superconducting phases with either $p$-wave or $(p+ip)$-wave order. 
In the following sections we will construct fully back reacted black hole solutions corresponding to these two phases. For orientation let us first recall the linearised static perturbations, which only involve the charged two-form $C$, that appear at the onset of the instability.

For the $p$-wave instability the key part of the perturbation is given by
\begin{align}
C&=\dots+c_{3}(r)\,dx_{1}\wedge[\sin\left(kx_{1}\right)\,dx_{2}+\cos\left(kx_{1}\right)\,dx_{3}]\,,
\end{align}
where the dots refer to terms that are determined from the function $c_3(r)$ via the equations of motion. 
The boundary conditions that are imposed on $c_3$ at the black hole horizon, located at $r=r_+$,
and at the $AdS$ boundary, located at $r\to \infty$, are, respectively,
\begin{align}\label{linbcs}
c_3(r)&\sim c_{3+}+\mathcal{O}(r-r_+)\,,\nn
c_3(r)&\sim c_{c_3}r^{-|m|}+\dots,\qquad r\to \infty\,.
\end{align}
The former ensures regularity at the black hole horizon and the latter ensures that the symmetry breaking is spontaneous.
Indeed the expectation value of the operator dual to $C$, with wave-number $k$, is proportional to $c_{c_3}$. 
Observe that when $k=0$, the perturbation is given by $C=\dots+c_{3}(r)\,dx_{1}\wedge dx_{3}$, corresponding to a $p$-wave vector order parameter pointing in the $-x_2$ direction (after taking a three-dimensional Hodge dual). This is also called
$-p_{x_2}$ (or $-p_{y}$) order.
This is
invariant under translations in three spatial directions and is also invariant under rotations
in the $(x_1,x_3)$ plane. When $k\ne 0$ the order parameter rotates in the $(x_2,x_3)$ plane as one moves along the $x_1$ direction
and there is a reduced helical (Bianchi VII$_0$) symmetry. Indeed there are still translations in the $x_2$ and $x_3$ directions, while
translations in the $x_1$ direction should be supplemented by a rotation in the $(x_2,x_3)$ plane.

For the $(p+ip)$-wave instability the key part of the perturbation is given by
\begin{align}
C&=\dots+e^{-i k x_1}ic_3(r) dx_1 \wedge (dx_2-i dx_3)\,,
\end{align}
where the dots again refer to terms that are determined from $c_3(r)$ from the equations of motion. 
The boundary conditions that are imposed on $c_3$ are the same as in \eqref{linbcs} with $c_{c_3}$
fixing the expectation value of the operator dual to $C$, with wave-number $k$.
When $k=0$, we have a superposition of the order parameter for $-p_{x_2}$ order and $i$ times the order parameter for
$p_{x_3}$ order, so this is also called $-p_{x_2}+ip_{x_3}$ order, or $p+ip$ for short. When $k=0$ 
we see that this is invariant under three spatial translations as well as a combination of rotations in the $(x_2,x_3)$ plane combined with a constant gauge transformation. 
When $k\ne 0$ we have, essentially, the same continuous symmetry; the only difference is that the translation
invariance in the $x_1$ direction needs to be supplemented with a constant gauge transformation. 

\begin{figure}
\centering
{\includegraphics[width=6cm]{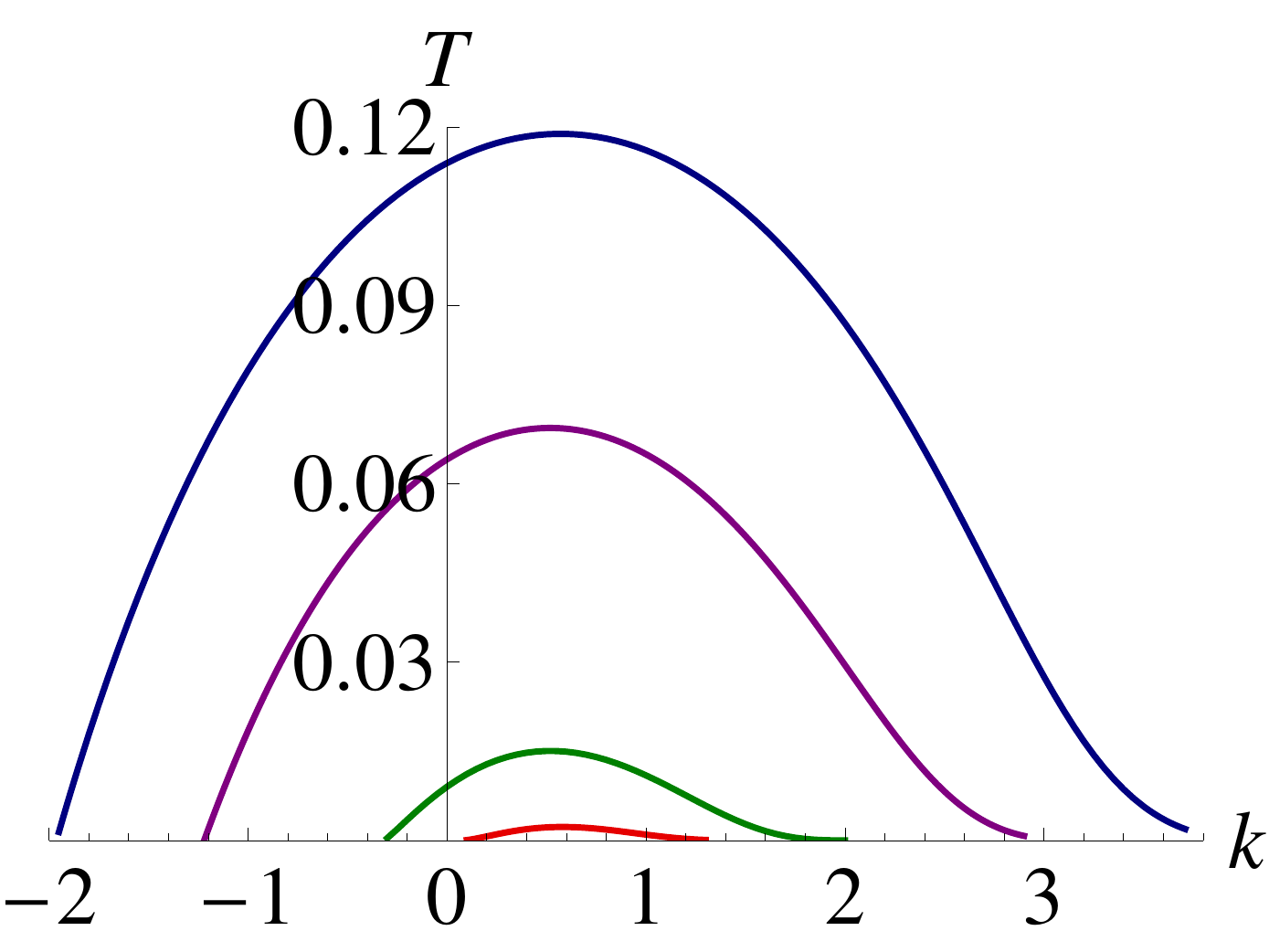}}
\caption{Plots of the critical temperature $T$ versus wave-number $k$ for the existence of normalisable, static perturbations of
the two-form about the AdS-RN black hole solution, for both $p$-wave and $(p+ip)$-wave order. The plots are for $m=2$ and, from top to
bottom $e=3.5$ (blue), $e=2.8$ (purple), $e=2$ (green) and $e=1.7$ (red). 
We have set the scale via $\mu=1$.}\label{figone}
\end{figure}

As shown in \cite{Donos:2011ff} the second order linear differential equation satisfied by $c_3(r)$ is exactly the same for both the $p$-wave and the $(p+ip)$-wave linearised perturbation. We can use the linearity of the equation to set $c_{3+}=1$ in the boundary conditions
given in \eqref{linbcs}. We then find that the critical temperature at which the
static perturbations exist is a function of wave-number $k$, giving\footnote{For fixed $(m,e)$, there can be additional curves of zero modes appearing at lower temperatures as noticed in other contexts e.g. \cite{Gubser:2008wv}. The zero modes corresponding to the higher critical temperature, as illustrated in figure \ref{figone}, are the ones that drive the instability towards the black holes of interest. The other zero modes correspond to branches of black holes which are expected to be unstable.} the characteristic ``bell curves" illustrated in figure \ref{figone}
for $m=2$ and several values of $e$. For each value of $k$ there will be two new branches of black hole solutions appearing at 
the temperature at which the static $p$-wave and $(p+ip)$-wave modes appear. 
This gives rise to two-parameter families of black hole solutions, labelled
by $k$ and $T$, for both the $p$-wave and the $(p+ip)$-wave cases. Constructing these solutions and finding the thermodynamically preferred solutions will be treated in subsequent sections. Observe that for fixed $m=2$, increasing the charge $e$ raises the critical temperature, as one expects for superconducting instabilities, and also broadens the width
of the curves, as was previously seen in \cite{Donos:2011ff}. Note also that for $m=2$ and small enough $e$, (for example $e=1.7$) 
the bell curves do not cross the $k=0$ axis.

\section{Helical $p$-wave black holes}\label{sectionp}
In this section we review the construction of \cite{Donos:2012gg}, expanding on some of the details, and obtain some new results.
The ansatz for the metric, gauge-field and two-form is given by
\begin{align}\label{eq:ansatzp}
ds^{2}&=-g\,f^{2}\,dt^{2}+g^{-1}{dr^{2}}+h^{2}\,\omega_{1}^{2}+r^{2}\,\left(e^{2\alpha}\,\omega_{2}^{2}+e^{-2\alpha}\,\omega_{3}^{2}\right)\,,\notag\\
A&=a\,dt\,,\nn
C&=(i\, c_{1}\,dt+c_2 dr)\wedge\omega_{2}+c_{3}\,\omega_{1}\wedge\omega_{3}\,,
\end{align}
where the one-forms $\omega_i$ are the left-invariant one-form of the
Bianchi type VII$_0$ Lie algebra
given by
\begin{align}\label{eq:one_forms}
&\omega_{1}=dx_{1}\,,\nn
&\omega_{2}=\cos\left(kx_{1}\right)\,dx_{2}-\sin\left(kx_{1}\right)\,dx_{3}\,,\nn
&\omega_{3}=\sin\left(kx_{1}\right)\,dx_{2}+\cos\left(kx_{1}\right)\,dx_{3}\,.
\end{align}
The ansatz depends on eight function $f$, $g$, $h$, $\alpha$, $c_{i}$ and $a$ which 
are all functions of the radial coordinate, $r$, and the wave-number, $k$, is a constant. Notice that the ansatz is periodic in the $x_1$ direction with period $2\pi/k$, a quantity also known as the pitch of the helix.
We are principally interested in the case that all of the coordinates $x_i$ 
are non-compact; the extension to the case of the coordinates
$x_i$ being compact is straightforward\footnote{In the compact case only discrete values of $k$ consistent with
the fixed period of $x_1$ are allowed. Furthermore, to obtain the thermodynamically preferred configurations, discussed below, 
one should
minimise the free energy as opposed to the free energy density.}. Notice that the ansatz is invariant under time translations associated with the Killing vector $\partial_t$ and the spacetime is static.
The ansatz is also spatially homogeneous with Bianchi $VII_0$ symmetry corresponding to the Killing vectors $\partial_{x_2}, \partial_{x_3}$, generating translations in the $x_2, x_3$ directions, 
and $\partial_{x_1}-k(x_2\partial_{x_3}-x_3\partial_{x_2})$, generating a helical motion consisting of
a simultaneous translation in the $x_1$ direction with a rotation in $(x_2,x_3)$ plane. The case of $k=0$ will be discussed below,
where we will see the appearance of an extra continuous symmetry.

The equations of motion we are interested in are obtained by substituting
the ansatz \eqref{eq:ansatzp} into the $D=5$ equations of motion \eqref{fulleom}. They can also be 
obtained by substituting the ansatz  \eqref{eq:ansatzp} directly into the $D=5$ action \eqref{eq:action} to obtain
\begin{align}
\label{eq:action1p}
S&=\int d^5x r^2hf\Bigg\{-g''-g'\left(\frac{3f'}{f}  +\frac{2 h'}{h} +\frac{4}{r} \right)+12\nn
&-2g\left[ \frac{f''}{f}+\frac{f'}{f}\left(\frac{2}{r}+\frac{h'}{h}\right)+\frac{h''}{h}+\frac{2h'}{r h}+\frac{1}{r^2}+\alpha'^2  \right]+\frac{a'^2}{2 f^2}-\frac{2 k^2}{h^2}\sinh(2\alpha)^2\nn
&-\frac{ c_3^2}{2 h^2 r^2}e^{2\alpha}+\frac{ c_1^2}{2 f^2 r^2 g}e^{-2\alpha}-\frac{gc_2^2}{2 r^2}e^{-2\alpha} \Bigg\}\nn
  &+\frac{1}{2m}\int d^5x\Bigg\{{c_1 c_ 3'}-{c_3 c_1'}+{2 e a c_2 c_3}+{2 k c_1
  c_ 2}\Bigg\}\,,
\end{align}
and then varying with respect to the eight functions, holding $k$ fixed. 
We find that $f$ and $g$ satisfy first order differential equations and that $h,\alpha,a$ and $c_3$ satisfy second order
equations with $c_1$ and $c_2$ determined via
\begin{align}\label{c1c2p}
c_{1}=-\frac{e^{2\alpha}}{e^{4\alpha}k^{2}+m^{2}\,h^{2}}\,\left( e^{2\alpha}kea c_{3}+mhfgc_{3}^{\prime}\right)\,,\notag\\
c_{2}=\frac{1}{fg}\frac{e^{2\alpha}}{e^{4\alpha}k^{2}+m^{2}h^{2}}\,\left( meahc_{3}-e^{2\alpha}kfgc_{3}^{\prime}\right)\,.
\end{align}
One can substitute these back into the action \eqref{eq:action1p} to obtain an action which will reproduce the equations of motion
for $f,g,h,\alpha,a$ and $c_3$.
The ansatz, and hence the equations of motion, are invariant under the following three scaling symmetries:
\begin{align}
\label{eq:symmetries1}
&r \to \lambda r\,, \quad (t,x_2,x_3) \to \lambda^{-1}(t,x_2,x_3)\,, \quad g\to \lambda^2 g\,, \quad a \to \lambda a\, ,\quad c_3 \to \lambda c_3\,; \nonumber\\
&x_1 \to \lambda^{-1} x_1\,, \quad h \to \lambda h\,,  \quad  k \to \lambda k\,, \quad c_3 \to \lambda c_3\,;\nonumber\\
&t \to \lambda t\,, \quad f \to \lambda^{-1}f\,, \quad a\to \lambda^{-1} a\,; \end{align}
where $\lambda$ is a constant.

\subsection{Asymptotic and near-horizon expansions}
We now discuss the boundary conditions to be imposed for the helical $p$-wave black holes solutions. 
As $r\to \infty$ we demand that we approach $AdS_5$ with asymptotic expansion
\begin{align}\label{uvexpp}
&g=r^{2}\,\left(1-{M}{r^{-4}}+\cdots \right),\qquad
f=f_{0}\left(1-{c_{h}}{r^{-4}}+\cdots\notag\right),\\
&h=r\,\left(1+{c_{h}}{r^{-4}}+\cdots \right),\qquad\,\,
\alpha={c_{\alpha}}{r^{-4}}+\cdots,\notag\\
&a=f_{0}\,\left(\mu+{q}{r^{-2}}+\cdots\right),\qquad
c_{3}={c_{c_3}}{r^{-\left|m\right|}}+\cdots,
\end{align}
which is specified by eight
parameters $M, f_0,c_h,c_\alpha,\mu,q,c_{c_3}$ and $k$. 
Notice that the boundary condition $h\sim r$ implies that the wave-number $k$ cannot be scaled away via \eqref{eq:symmetries1}.
However, the scaling symmetries do allow us to set $\mu=f_0=1$, and we will do so
later (it is helpful to keep them to discuss the thermodynamics). 
Observe that the fall-off of $c_3$ is chosen so that the charged operator dual to the two-form 
$C$ has no deformation but can acquire, {\it spontaneously}, an expectation value proportional to $c_{c_3}$ 
which is spatially modulated in the $x_1$ direction with period $2\pi/k$. The holographic interpretation of the other UV parameters will be given below.

We also demand that we have a regular black hole event horizon located at $r=r_+$.
As $r\to r_+$, the functions have the analytic expansion
\begin{align}
\label{eq:IRexpp}
g&=g_+(r-r_+)+\cdots\,,\qquad\,\,
f=f_++\cdots\,,\nn
h&=h_++\cdots\,,\qquad\qquad\qquad
\alpha=\alpha_++\cdots\,,\nn
a&=a_+(r-r_+)+\cdots\,,\qquad
c_3=c_{3+}+\cdots\,.
\end{align} 
We find that the full IR expansion is fixed in terms of the six constants $f_+, h_+,\alpha_+, a_+,c_{3+}$ and $r_+$. In particular, the coefficient $g_+$ is fixed by these constants:
\begin{equation}
g_+= r_+(4-\frac{a_+^2}{6 f_+^2})-\frac{c_{3+}^2e^{2\alpha_+}}{12r_+h_+^2}\,.
\end{equation}

The equations of motion give four second order differential equations for $h,\alpha,a,c_3$ and 
two first order equations for $g,f$ and hence a solution is specified by ten integrations constants. On the other hand we have fourteen parameters in the boundary conditions minus two for the scaling symmetries. We thus expect a two-parameter family of black hole solutions
that can be specified by temperature $T$ and wave number $k$.

\subsection{Numerical solutions}
\begin{figure}
\centering
\subfloat[]{\includegraphics[width=5cm]{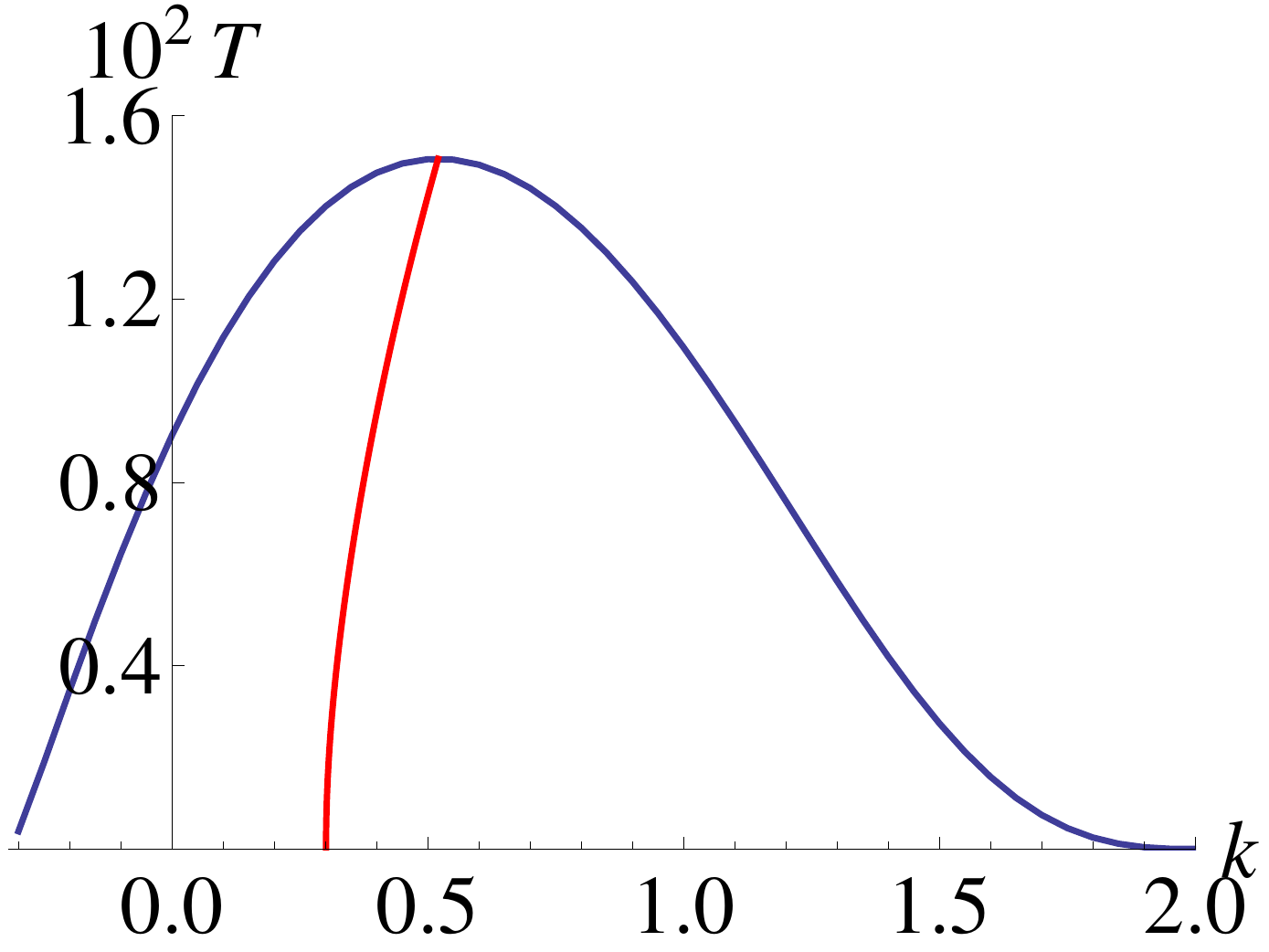}}\hskip 4 em
\subfloat[]{\includegraphics[width=5cm]{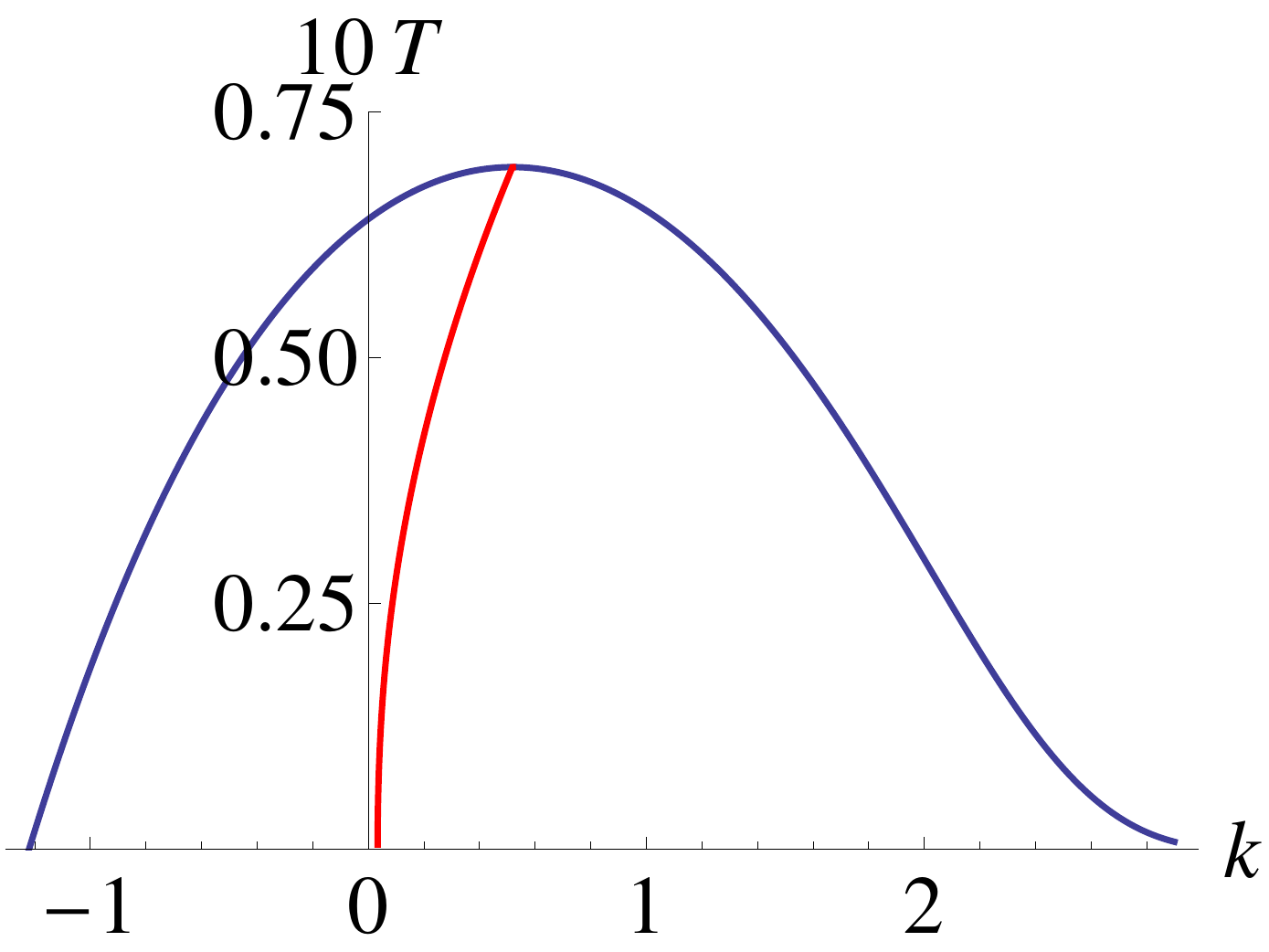}}\\
\subfloat[]{\includegraphics[width=5cm]{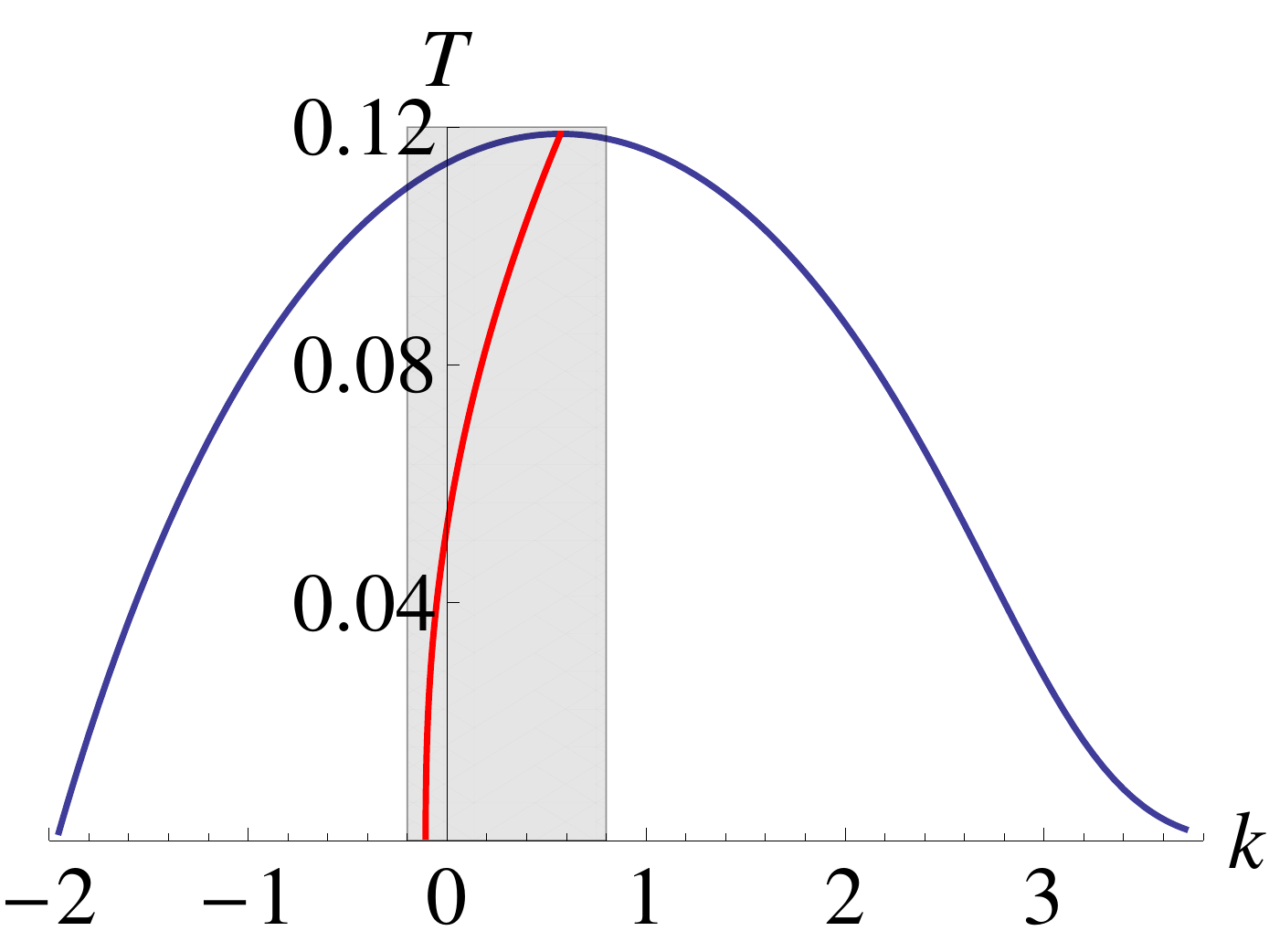}}\hskip 3 em
\subfloat[]{\includegraphics[width=6cm]{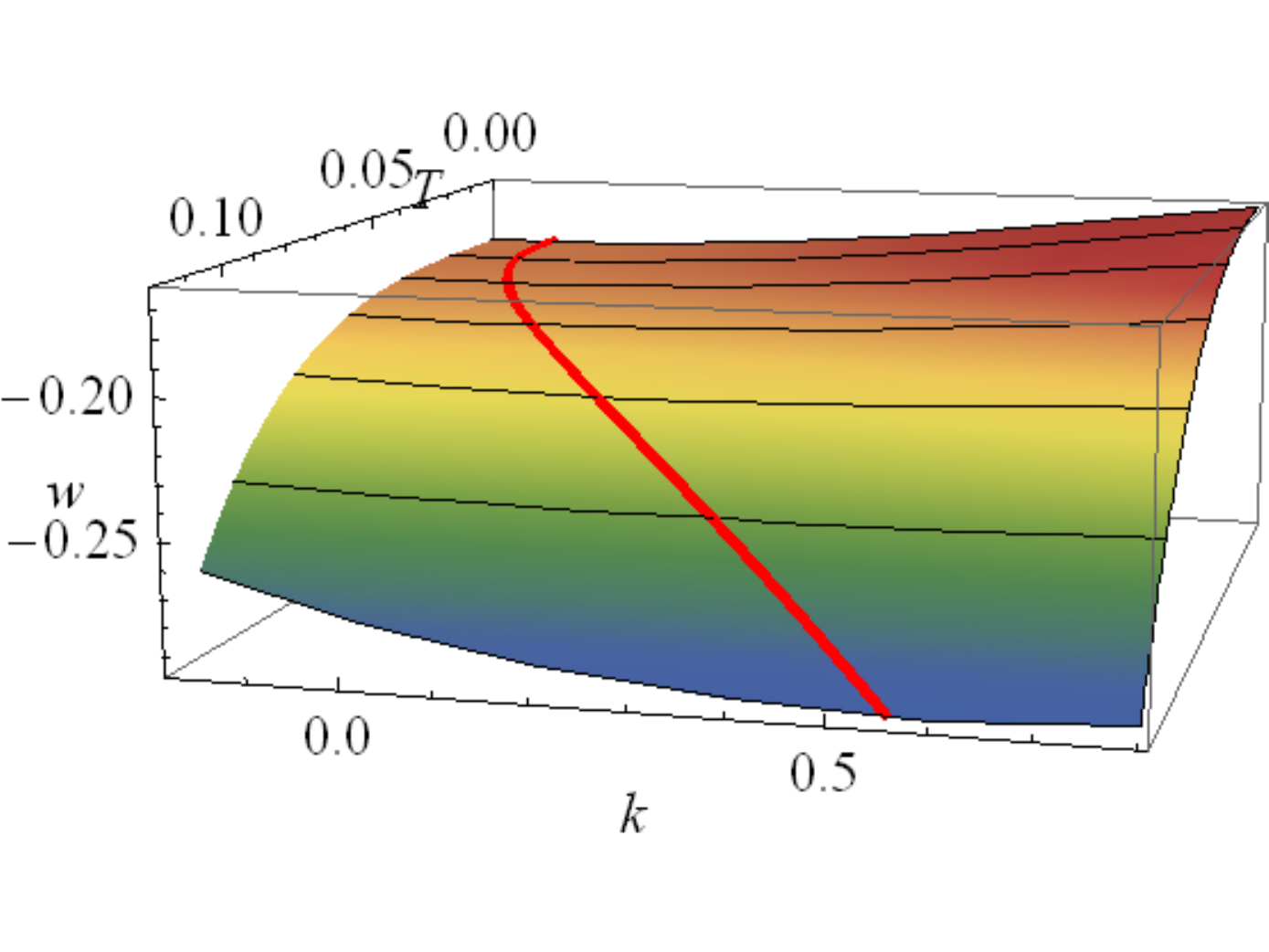}}
\caption{$p$-wave black holes for $m=2$ and $e=2$ (panel (a)), $e=2.8$ (panel (b)) and $e=3.5$ (panels (c) and (d)). 
Each point under the bell curve corresponds to a $p$-wave black hole
at temperature $T$ and wave-number $k$. 
All of these black holes have smaller free energy than the AdS-RN black hole at the same temperature.
The red curve in each panel corresponds to the thermodynamically preferred branch of black holes
that minimise the free energy density with respect to $k$ at fixed $T$. 
In panel (d) we have plotted the free-energy density as a function of $T$ and $k$ for
the representative case of $e=3.5$; the slice of $k$-values is represented by the grey shaded region in panel (c).
For $e=2$ we see in panel (a) that the pitch ($2\pi /k$) monotonically increases to a constant positive value
at $T=0$. For larger values of $e$, such as $e=3.5$ in panel (c), the black holes exhibit pitch inversion, with $k=0$ at some non-zero 
$T$. We have set $\mu=1$.}\label{figtwo}
\end{figure}

\begin{figure}
\centering
\subfloat[]{\includegraphics[width=4.5cm]{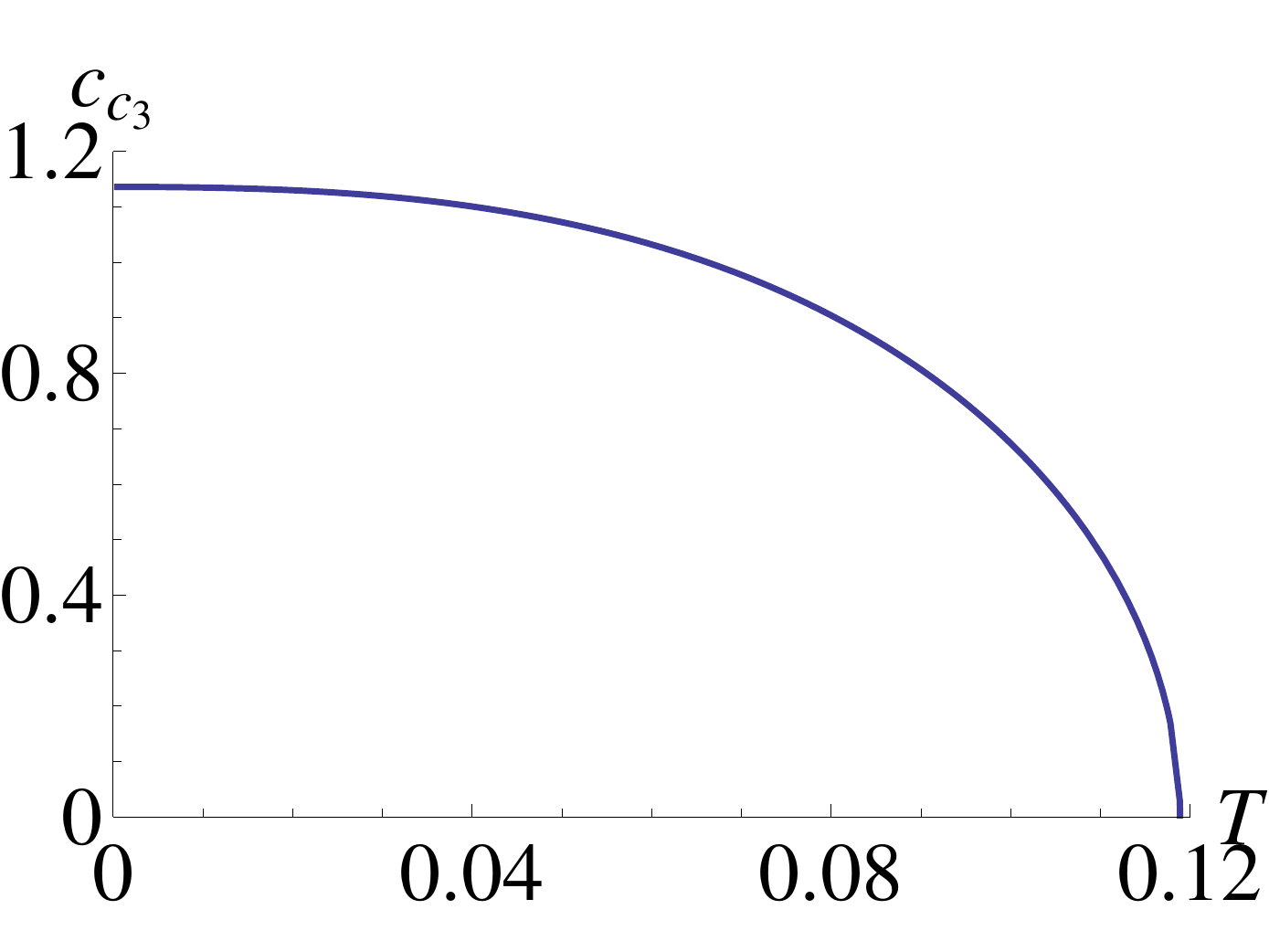}}\hskip 1 em
\subfloat[]{\includegraphics[width=4.5cm]{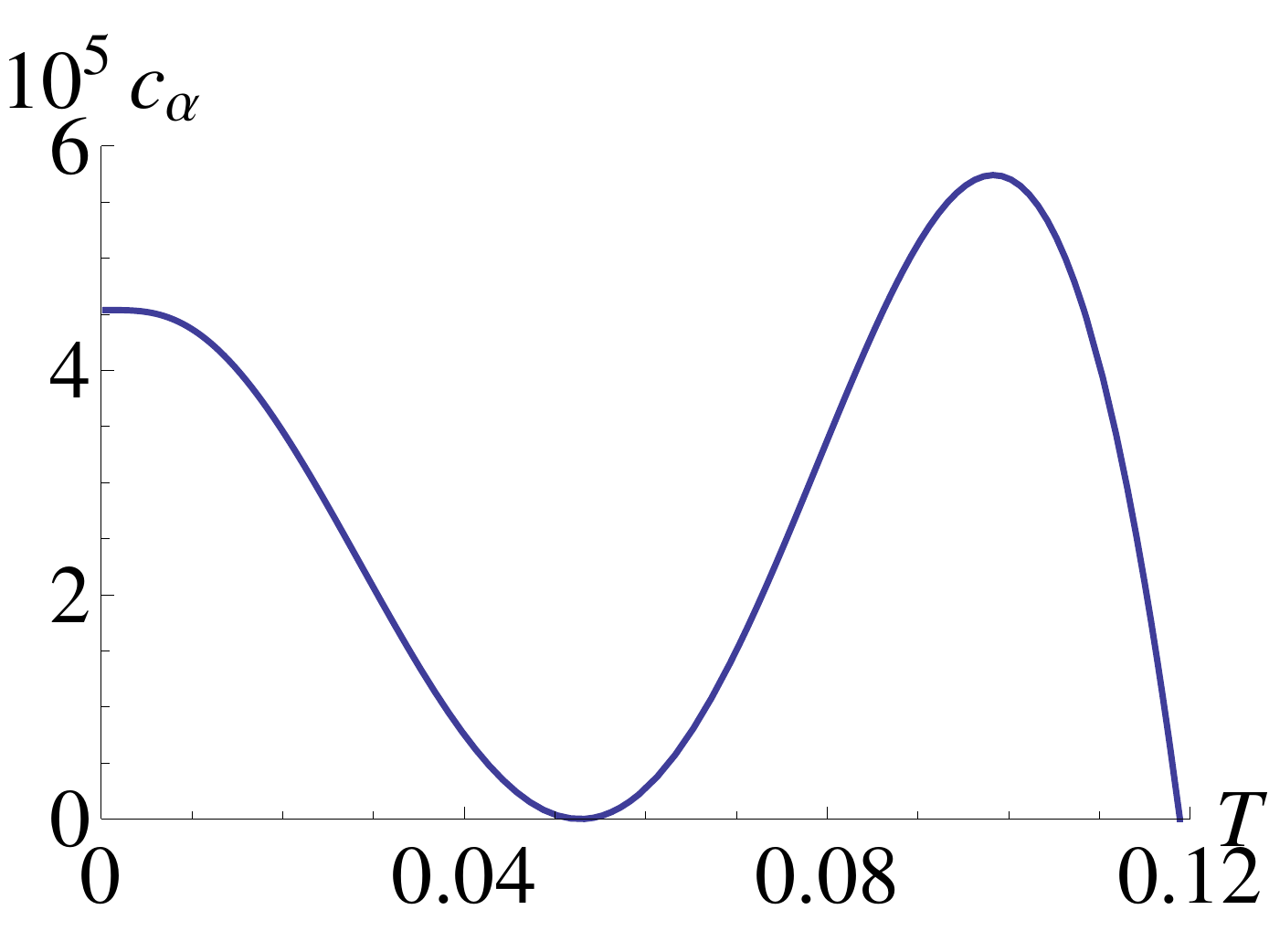}}\hskip 1 em
\subfloat[]{\includegraphics[width=4.5cm]{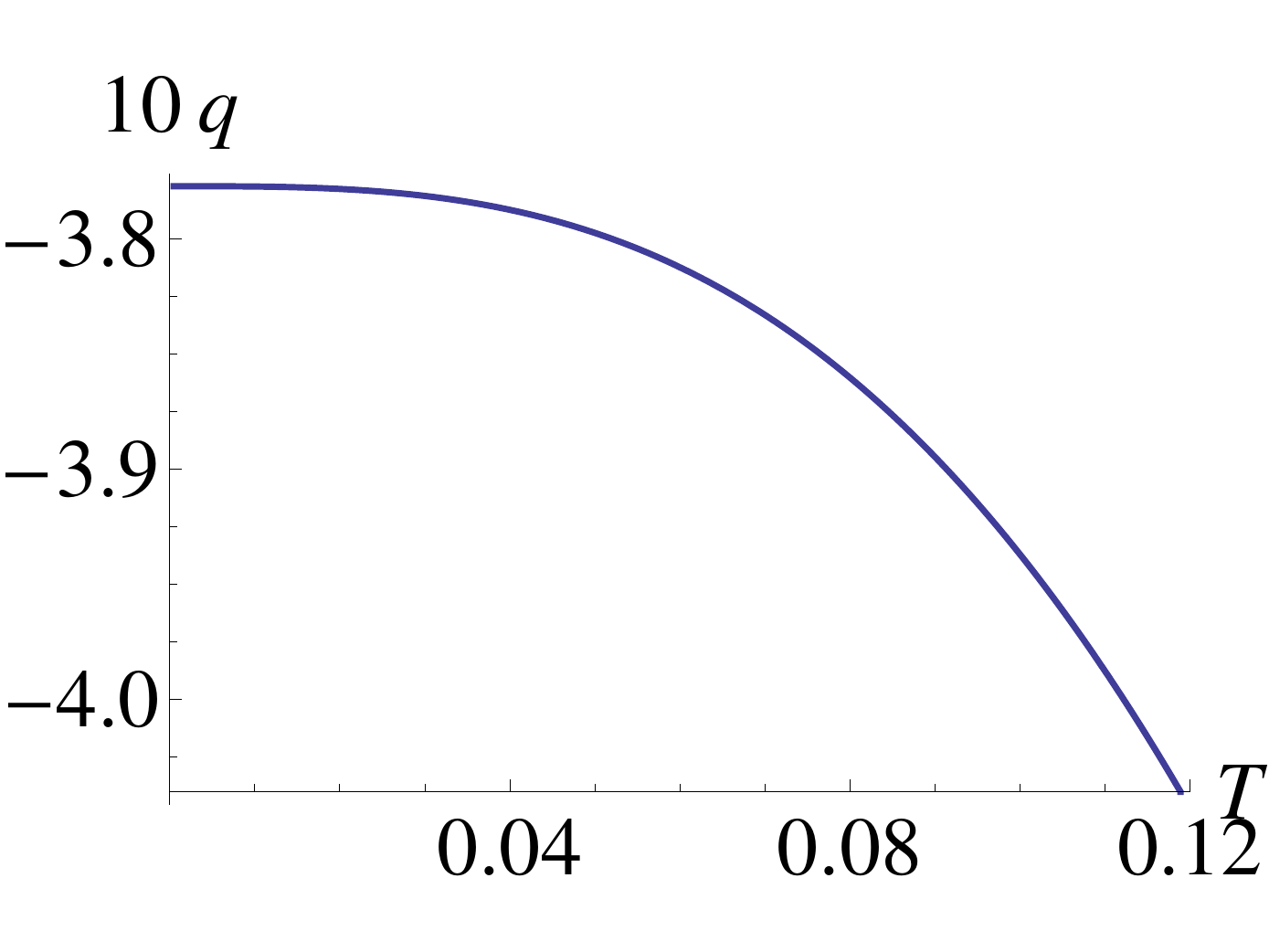}}
\caption{Properties of $p$-wave black holes for $m=2$ and $e=3.5$ as a function of $T$ for the
thermodynamically preferred branch (the red lines in figures \ref{figtwo}(c) and \ref{figtwo}(d)).
Panel (a) plots $c_{c3}$ which fixes the $p$-wave order parameter; panel (b) plots $c_\alpha$ which fixes spatial modulation 
of the stress tensor - observe that it goes to zero at the temperature where the pitch inversion occurs; 
panel (c) plots $q$ which fixes the charge density.
Note that the thermodynamically preferred black holes have $c_h=0$.
We have set $\mu=1$.}\label{figthree}
\end{figure}

We have numerically constructed these black holes and have summarised some of the results in figures \ref{figtwo} and \ref{figthree}, for
$m=2$ and various values of $e$ (note that black holes with $m=1.7$ and $e=1.88$ were constructed in \cite{Donos:2011ff}).
Various aspects of these black holes, including their thermodynamics and ground states, will be discussed in the
following subsections. In figure \ref{figtwo} we display the two-parameter family of $p$-wave black holes, corresponding to
the bell curves in figure \ref{figone}, including the thermodynamically preferred branch obtained by minimising the free-energy density
with respect to $k$ at fixed $T$, as we discuss below. Note that all black holes have smaller free energy than the AdS-RN black hole and
that the transition to the $p$-wave preferred branch is second order.
In figure \ref{figthree} we have plotted various physical quantities for the preferred branch for the representative case of
$m=2$, $e=3.5$; other values of $e$ are similar. We discuss the behaviour of solutions as $T\to0$ in section \ref{tzerolimp}; in all cases it appears that
the black holes approach zero entropy ground states with a
behaviour consistent with an emergent scaling symmetry.

\subsection{Thermodynamics}\label{thermp}
We analytically continue by setting $t=-i\tau$. Regularity of the solution at $r=r_+$ is achieved by making 
 $\tau$ periodic with period $\Delta \tau=4 \pi /(g_+f_+)$ corresponding to temperature $T=(f_0 \Delta \tau)^{-1}$. Note that to ensure we get
a real gauge-field and two-form field near $r=r_+$ we should set $a_+=i \bar{a}_+$ and $c_{1+}=i \bar{c_1}_+$.
We can also read off the area of the event horizon  and since we are working in units  with $ 16 \pi G=1$, we deduce that entropy density is given by 
\begin{equation}
s=4 \pi r_+^2 h_+\,.
\end{equation}
We will consider the total Euclidean action, $I_{Tot}$, defined as
\begin{equation}
I_{Tot}=I+I_{ct}\,,
\end{equation}
where $I=-iS$ and the counter-term action is given by an integral on the boundary $r\to \infty$:
\begin{equation}\label{ctermp}
I_{ct}=\int d\tau d^3 x \sqrt{-g_\infty}(-2 K+6+ \cdots)\,.
\end{equation}
Here $K=g^{\mu \nu} \nabla_\mu n_\nu$ is the trace of the extrinsic curvature of the boundary, where $n^\mu$ is an outward pointing normal unit vector, and $g_\infty$ is the determinant of the induced metric. The ellipsis refers to terms which will not be relevant for the ansatz and boundary conditions that we are considering. For our ansatz we have
\begin{equation}\label{ctermpex}
I_{ct}=Vol_3 \Delta \tau \lim_{r \to \infty} r^2 h f g^{1/2}[6-2g^{1/2} (\frac{2}{r} +\frac{f'}{f}+ \frac{h'}{h})- g^{-1/2} g']\,,
\end{equation}
where $Vol_3=\int dx_1 dx_2 dx_3$. We next point out two equivalent ways to write the bulk part of the Euclidean action on-shell:
\begin{align}\label{act2waysp}
I_{OS}=& Vol_3 \Delta\tau \int_{r_+}^\infty \left(2 r g h f\right)'\,,\nonumber \\
          =& Vol_3 \Delta\tau \int_{r_+}^\infty \left(r^{2}hfg^{\prime}+2r^{2}hgf^{\prime}-\frac{r^{2}haa^{\prime}}{f}
          +\frac{c_1 c_3}{2m}\right)'\,,
        \end{align}
where $c_1$ is given in \eqref{c1c2p}.
Notice that the first expression only receives contributions from the boundary at $r \to \infty$ since $g(r_+)=0$, while the second expression also receives contributions from $r=r_+$. 
We next define the free energy $W=T[I_{Tot}]_{OS}\equiv w Vol_3$.
Using the UV and the IR expansions \eqref{uvexpp}, \eqref{eq:IRexpp}
we obtain the following expression for the free energy density:
\begin{align}
\label{eq:OSactionp}
w&= -M\, ,\nonumber \\
          &= 3M +8 c_h+2 \mu q -s T\,,
\end{align}
and hence the Smarr-type formula:
\begin{align}\label{smarrp}
4M +8 c_h+2 \mu q -s T=0\,.
\end{align}

An on-shell variation of the total action $[I_{Tot}]_{OS}$, for fixed $k$, gives
\begin{equation}
[\delta I_{Tot}]_{OS}=Vol_3 \Delta \tau [\delta f_0(3M +8 c_h +2 \mu q )+2 f_0 q \delta \mu]\,.
\end{equation}
In this variation we are holding $\Delta \tau$ fixed and hence $\Delta \tau \delta f_0=-T^{-2} \delta T$. 
We thus deduce that $w=w(T, \mu)$ and the first law
\begin{equation}\label{firstlawp}
\delta w=-s \delta T+2 q \delta \mu\,.
\end{equation}

We now compute the expectation value of the boundary stress-energy tensor. 
The relevant terms are given by \cite{Balasubramanian:1999re}
\begin{equation}\label{stressy}
\langle T_{\mu\nu}\rangle=\lim_{r\to\infty} r^2 [-2 K_{\mu\nu}+2(K-3)g_{\infty \mu\nu}+\cdots]\,.
\end{equation}
Using the asymptotic expansion \eqref{uvexpp}, we obtain, after setting $f_0=1$,
\begin{align}
\label{eq:bdySTp}
\langle{T_{tt}}\rangle&=3 M +8 c_h\,,\nonumber\\
\langle{T_{x_1 x_1}}\rangle&=M+8 c_h\,,\nonumber\\
\langle{T_{x_2 x_2}}\rangle&=M+8c_\alpha\cos(2kx_1)\,,\nonumber\\
\langle{T_{x_3 x_3}}\rangle&=M-8c_\alpha\cos(2kx_1)\,,\nn
\langle{T_{x_2 x_3}}\rangle&=-8c_\alpha\sin(2kx_1)\,.
\end{align} 
We easily see that this is traceless with respect to the flat boundary metric.
We also note that defining the energy density $\varepsilon=3M +8 c_h$, we can rewrite $w=\varepsilon-sT +2 \mu q$ and
the first law takes the form $\delta \varepsilon=T \delta s -2 \mu \delta q$.

The next step is to calculate the expectation value of the current. The relevant terms are given by
\begin{equation}\label{jterms}
\langle J_\mu\rangle =-\lim_{r \to \infty} r^3[F_{r \mu}+\cdots]\,,
\end{equation}
where the ellipsis refers to terms that will not be relevant here. Using \eqref{uvexpp}, we find that the only non-zero component is
given by
\begin{align}
&\langle{J}_{t}\rangle=2 q\,,
\end{align}
where we have again set $f_0=1$. As expected, $q$ fixes the charge density.

\subsubsection{Variation of $w$ with respect to $k$}
We have constructed two-parameter families of $p$-wave black hole
solutions that can be labelled by temperature $T$ and
wave-number $k$ (see figure \ref{figtwo}(d)). 
For a given temperature we are interested in the solution labelled by $k_{min}$ which minimises
the free energy density (as in figures \ref{figtwo}(a)-(c)). These black holes are specified by varying the action $I_{Tot}$ with respect to $k$ and setting
it to zero on-shell. For the numerically constructed black holes of \cite{Donos:2012gg} it was shown that
these solutions have $c_h=0$. Here we will directly prove this fact, which also
follows from the general results for periodic black branes obtained in \cite{Donos:2013cka}, as we will explain.

The variation of the Euclidean action $I_{Tot}$ with respect to $k$, which only gets contributions from the bulk piece $I$, gives
\begin{align}\label{varabc}
k\,\partial_{k}I_{Tot}=Vol_3\Delta\tau \int_{r_+}^\infty dr \left[-\frac{k}{m}\,c_{1}c_{2}+\frac{4k^{2}r^{2}f}{h}\,\sinh^{2}(2\alpha)\right]\,,
\end{align}
where $c_1, c_2$ are given in \eqref{c1c2p}.
After imposing the equations of motion we find that the integrand can be rewritten as a total derivative leading to
\begin{align}
&[k\,\partial_{k}I_{Tot}]_{OS}=\nn
&I_{OS}+Vol_3\Delta\tau\int_{r_+}^\infty dr \left[  - 2r^{2}h^{\prime}fg-\frac{1}{2}\frac{e^{2\alpha}hfg}{e^{4\alpha}k^{2}+m^{2}h^{2}}c_{3}c_{3}^{\prime}-\frac{ke}{2m}\,\frac{e^{4\alpha}a}{e^{4\alpha}k^{2}+m^{2}h^{2}}c_{3}^{2}          \right]'\,,
\end{align}
where $I_{OS}$ was given in \eqref{act2waysp}. Evaluating the two terms on the right hand side by substituting the UV and the IR expansions \eqref{uvexpp}, \eqref{eq:IRexpp}, 
we find that the divergent pieces cancel leading to the finite result $[k\,\partial_{k}I_{Tot}]_{OS}=Vol_3 T^{-1}8c_h$.  
Hence, at constant $T$ we have 
\begin{align}\label{kvarp}
k\partial_k w=8c_h\,,
\end{align}
and 
hence the one-parameter family of thermodynamically preferred black holes
satisfy the necessary condition
\begin{align}
\partial_k w=0\quad\Rightarrow\quad c_h=0\,,
\end{align}
We will see in section \ref{solkzerop} that this is {\it not} a sufficient condition for picking out the preferred branch. 
The thermodynamically preferred black holes are labelled by the red lines in figure \ref{figtwo}.
Note that for this one-parameter family of black holes (labelled by the temperature) the stress tensor of the boundary CFT given in \eqref{eq:bdySTp} is still spatially modulated in the $(x_2,x_3)$ plane.

\subsubsection{Connection with the results of \cite{Donos:2013cka}}\label{conp}
A general analysis of the thermodynamics of periodic black branes was carried out in 
\cite{Donos:2013cka}. It was shown that several results can be immediately obtained from the boundary stress tensor $T^{\mu\nu}$
and the current $J^\mu$. Specifically, in the present set up it was shown\footnote{Note that we have used the fact that there are no source terms for the
two-form $C$.} that
\begin{align}\label{eq:final_var_illp}
w&=-Ts-\bar J^{t}\mu+\bar T^{tt}\,,\nn
w&=-\bar T^{x_2x_2}=-\bar T^{x_3x_3}\,\nn
\delta w&=-\bar J^{t}\delta \mu-s{\delta T}+\frac{\delta k}{k}\left(w+T^{x_1x_1}\right)\,,
\end{align}
where the bars refer to quantities averaged in the $x_1$ direction;
$\bar J^t=(k/2\pi)\int_0^{2\pi/k} dx_1 J^t$ and $\bar T^{x_ix_i}=(k/2\pi)\int_0^{2\pi/k} dx_1 T^{x_ix_i}$. Note also that we have used the conservation of the
stress tensor to deduce that $T^{x_1x_1}$ is constant.
Using the expressions for the stress tensor and current given in \eqref{eq:bdySTp}, \eqref{jterms} and substituting into \eqref{eq:final_var_illp} we find that
\begin{align}\label{eq:final_var_ill}
w&=-Ts+2q\mu+3M+8c_h\,,\nn
w&=-M\,,\nn
\delta w&=2q\delta \mu-s{\delta T}+\frac{\delta k}{k}\left(w+M+8c_h\right)\,.
\end{align}
We thus recover the two expressions for the action derived earlier \eqref{eq:OSactionp} and hence the Smarr formula \eqref{smarrp}.
We also obtain the first law given in \eqref{firstlawp}, \eqref{kvarp}. In particular, for the case that $x_1$ is non-compact, 
we should impose $\delta w/\delta k=0$ and we thus have $c_h=0$ for the thermodynamically preferred black holes
and furthermore that $T^{x_1x_1}=\bar T^{x_2x_2}=\bar T^{x_3x_3}$ as also noted in \cite{Donos:2013cka}.

\subsection{Solutions with $k=0$}\label{solkzerop}
For values of $m,e$ where the bell curves cross the $k=0$ axis (see figures \ref{figone} and \ref{figtwo}),
there are $p$-wave black hole solutions with $k=0$. These black holes are not thermodynamically preferred,
except at the specific temperature at which the red branch of thermodynamically preferred solutions
intersects the $k=0$ line, corresponding to pitch inversion, as in figure \ref{figtwo}(c) for $m=2$ and $e=3.5$. 
For $m=2$ and $e\sim 2.9$, the $k=0$ black
hole is the $T=0$ ground state of the $p$-wave black holes. In this subsection we discuss the entire $k=0$ branch.

The black holes with $k=0$ are translationally invariant in the $x_1$ direction. We can use the results of \cite{Donos:2013cka}  to 
conclude\footnote{ For these solutions we can define the free energy density in some finite interval $x_1\in [0,L]$
and then take the limit $L\to\infty$. The variation of the free energy density with respect to varying $L$ is given by
$-(w+M+8c_h)$ \cite{Donos:2013cka} and the translation invariance in the $x_1$ direction implies that this vanishes.}
that for this branch there will be a Smarr formula which implies $w+M+8c_h=0$. Since we also have 
$w=-M$ from \eqref{eq:OSactionp} we deduce that $c_h=0$. 
In addition, we have $\bar T^{x_ix_i}=M=T^{x_1x_1}$ for $i=2,3$. On the other hand
from \eqref{eq:bdySTp} we now have $\bar T^{x_ix_i}=M+c_\alpha$ for $i=2,3$ 
and hence we conclude that for the $k=0$ black holes we also have $c_\alpha=0$. 

By analysing the equations of motion near $r\to \infty$ one can see that $c_h=c_\alpha=0$
implies that functionally we have $h=re^{-\alpha}$. Indeed one can show that setting $h=re^{-\alpha}$ 
is a consistent truncation of the
equations of motion corresponding to the following ansatz  for the $k=0$ $p$-wave black holes:
\begin{align}\label{eq:ansatzpk=0}
ds^{2}&=-g\,f^{2}\,dt^{2}+g^{-1}{dr^{2}}+r^{2}e^{-2\alpha}(dx_1^2+dx_3^2)+r^2e^{2\alpha}\,dx_{2}^{2}\,,\notag\\
A&=a\,dt\,,\nn
C&=(i\, c_{1}\,dt+c_2 dr)\wedge dx_{2}+c_{3}\,dx_{1}\wedge dx_{3}\,,
\end{align}
and now with
\begin{align}\label{c1c2pt}
c_{1}=-\frac{e^{3\alpha}fgc_{3}^{\prime}}{mr}\,\,,\qquad
c_{2}=\frac{e^{3\alpha} eac_{3}}{mfgr}\,.
\end{align}
Notice that $c_3$ in the two-form picks out the $x_2$ direction as being preferred. The metric ansatz reflects
this anisotropic structure, but the specific black hole solutions have $c_\alpha=0$ leading to an isotropic
energy-momentum tensor (see \eqref{eq:bdySTp}). We also note that these $k=0$ black holes have more symmetry than those with $k\ne 0$: the
Bianchi VII$_0$ symmetry is replaced with three spatial translations as well as rotations in the $(x_1,x_3)$ plane and this
is related to the fact that $h=re^{-\alpha}$ is a consistent truncation of the equations of motion.

\begin{figure}
\centering
\subfloat[]{\includegraphics[width=5cm]{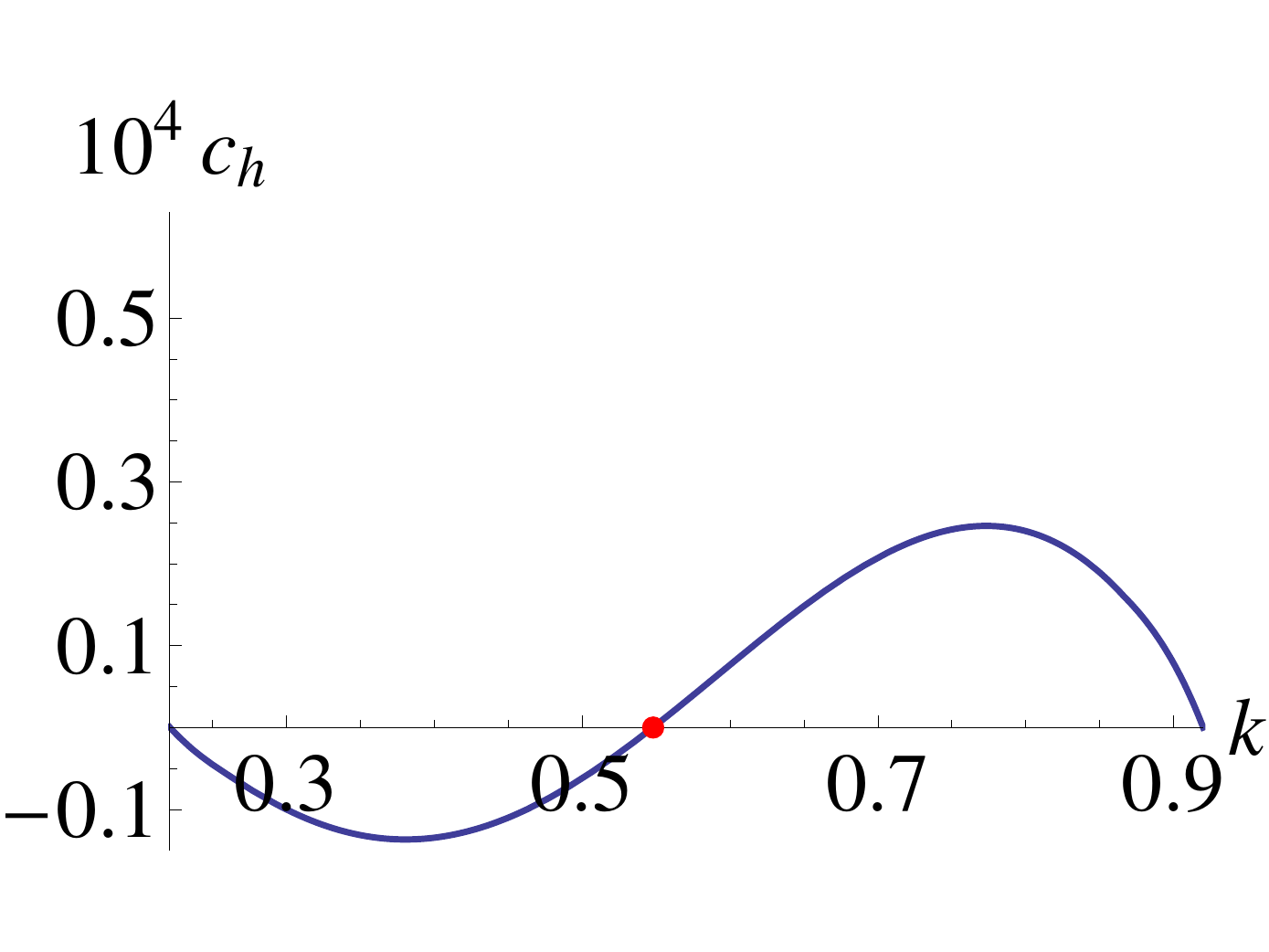}}\hskip 2 em
\subfloat[]{\includegraphics[width=5cm]{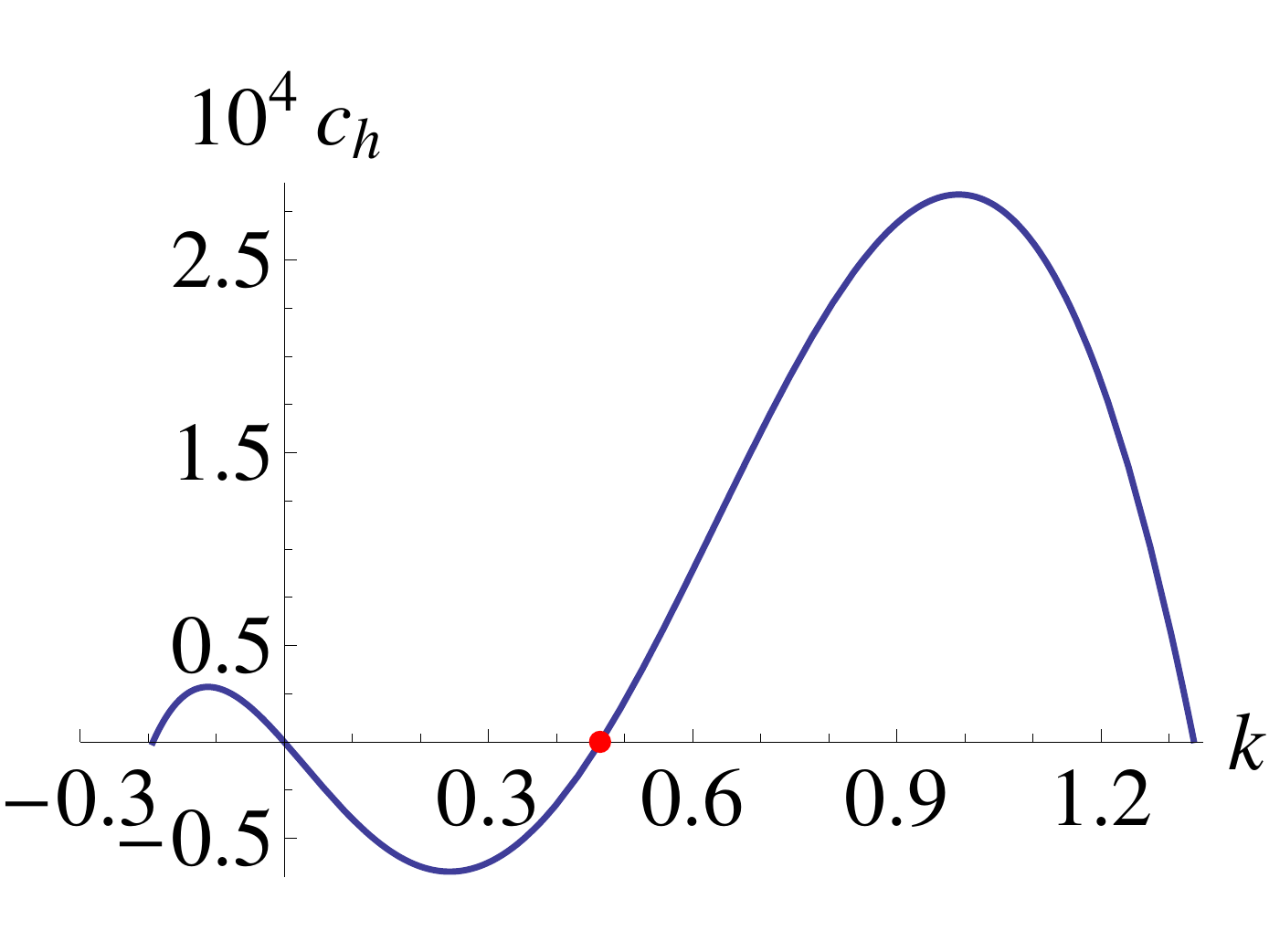}}
\caption{A plot of $c_h$ versus $k$ for $p$-wave black holes with $m=2$ and $e=3.5$ at temperature $T=0.117$ 
(panel (a)) and $T=0.11$ (panel (b)). The range of $k$ plotted corresponds to the range of $k$ in the bell curve in figure
\ref{figtwo}(c) for the two temperatures. In panel (a) we see that there is one zero of $c_h$ in the interior, 
corresponding to the red branch, while in panel (b) there are two zeroes, one corresponding to the red branch (marked with
the red dot) and the other
to the $k=0$ branch. We have set the scale via $\mu=1$.}\label{figfour}
\end{figure}

It is helpful at this point to address a potential confusion. We have argued that the thermodynamically preferred black holes
which minimise the free energy with respect to $k$ have $c_h=0$. The converse is not true. For example, the $k=0$ branch
of black holes has $c_h=0$ but they do not (in general) comprise an extrema of the free-energy density. It is 
illuminating to plot the behaviour of $c_h$ against $k$ for some representative temperatures, as in figure \ref{figfour}.
It is also interesting to observe that from \eqref{varabc} we can conclude that at $k=0$ we have
$\partial_kw=-1/m\int_{r_+}^\infty dr c_1 c_2$. In particular, we can establish that the $k=0$ solutions
are unstable by obtaining a non-vanishing result for the integral for the 
$k=0$ branch of solutions. Furthermore, by taking a derivative of \eqref{kvarp} with respect to $k$, we deduce
that, with non-diverging $\partial_k^2w$, this integral is given by $8\partial_k c_h$ at $k=0$.

\subsection{Pitch inversion}

In figure \ref{figtwo} we see that the value of $k$ for the thermodynamically preferred black holes
monotonically decreases as the temperature is lowered, for $m=2$ and various values of $e$. 
When $m=2$ and $e=2$, for example, $k$ decreases to a value $k>0$ at $T=0$. 
As $e$ is increased we observe the phenomena of pitch inversion. 
For example, when $m=2$ and $e=3.5$ the pitch ($2\pi/k$) first increases, becoming infinite (i.e. $k=0$) at 
$T\approx 0.053$ and 
then changes sign and decreases in magnitude leading to a value $k<0$ at $T=0$.  
It is interesting to note that precisely at the pitch inversion temperature the symmetry of the black hole solutions is enhanced.

Notice that for $m$=2 the numerics indicate that  if $e\lesssim 2.9$ then the preferred black holes hit a $T=0$ ground state with $k>0$, while if 
$e\gtrsim2.9$ they hit a $k<0$ ground state. For the intermediate value\footnote{It is difficult, numerically, to extract the precise value of $e$ at which this occurs.}, 
$e\sim 2.9$, the ground state has $k=0$. We will describe these ground states and their infrared scaling behaviour in the next subsections.

\subsection{Scaling symmetry ground states with $k\ne 0$ and $k=0$}\label{scalp}
In this subsection we present some solutions to the equations of motion with scaling symmetry. We also discuss the
corresponding domain wall solutions which interpolate between $AdS_5$ in the UV and the scaling solutions in the IR 
which provide putative ground states for the $p$-wave black holes at $T=0$.

There are scaling solutions with $k\ne 0$ and Bianchi VII$_0$ symmetry that were 
first discussed in \cite{Donos:2012gg} (and are similar to those in \cite{Iizuka:2012iv}).
We will see that these appear to correspond to the IR behaviour of the $T=0$ $p$-wave black holes
which have $k>0$ at $T=0$ (i.e. $e\lesssim 2.9$). 
There are also scaling solutions with $k=0$ that are new and share some similarities to the scaling solutions
discussed in \cite{Taylor:2008tg}. These appear to correspond to the IR behaviour of the $T=0$ $p$-wave black holes
which have $k=0$ at $T=0$ (i.e. $e\sim 2.9$). Interestingly, the $p$-wave black holes which have $k<0$ at $T=0$ (i.e. $e\gtrsim 2.9$) appear to approach $AdS_5$ in
the IR. We will explain how the corresponding domain wall solutions can be constructed using $k$-dependent relevant 
perturbations in the IR.

\subsubsection{Scaling solutions with $k\ne 0$ and Bianchi VII$_0$ symmetry}\label{scalhelp}
When $k\ne 0$ the equations of motion admit scaling solutions of the form \cite{Donos:2012gg} 
\begin{align}\label{helfp}
g=L^{-2}r^2,\quad f=\bar f_0r^{z-1},\quad h=k h_0,\quad \alpha=\alpha_0,\quad
a=\bar f_0a_0 r^z,\quad c_3=k c_3^0r\,,
\end{align}
where $z, L,\bar f_0, h_0,\alpha_0,a_0$ and $c_3^0$ are constants.
For example, the metric for these solutions reads
\begin{align}\label{helfpmet}
ds^{2}&=-(\bar f_0^2L^{-2})r^{2z}dt^{2}+L^2\frac{dr^{2}}{r^2}+(k^2h_0^{2})dx_{1}^{2}+r^{2}\,\left(e^{2\alpha_0}\,\omega_{2}^{2}+e^{-2\alpha_0}\,\omega_{3}^{2}\right)\,,
\end{align}
where $\omega_i$ are the left invariant Bianchi VII$_0$ one-forms given in \eqref{eq:one_forms}.
By scaling $t$ and $x_1$ we can set $\bar f_0=k=1$. These fixed point solutions are
invariant under the anisotropic scaling 
\begin{align}\label{knzsc}
r\to \lambda^{-1}r,\quad t\to\lambda^z t,\quad x_{2,3}\to \lambda x_{2,3},\quad x_1\to x_1\,.
\end{align} 

\begin{figure}
\centering
{\includegraphics[width=6cm]{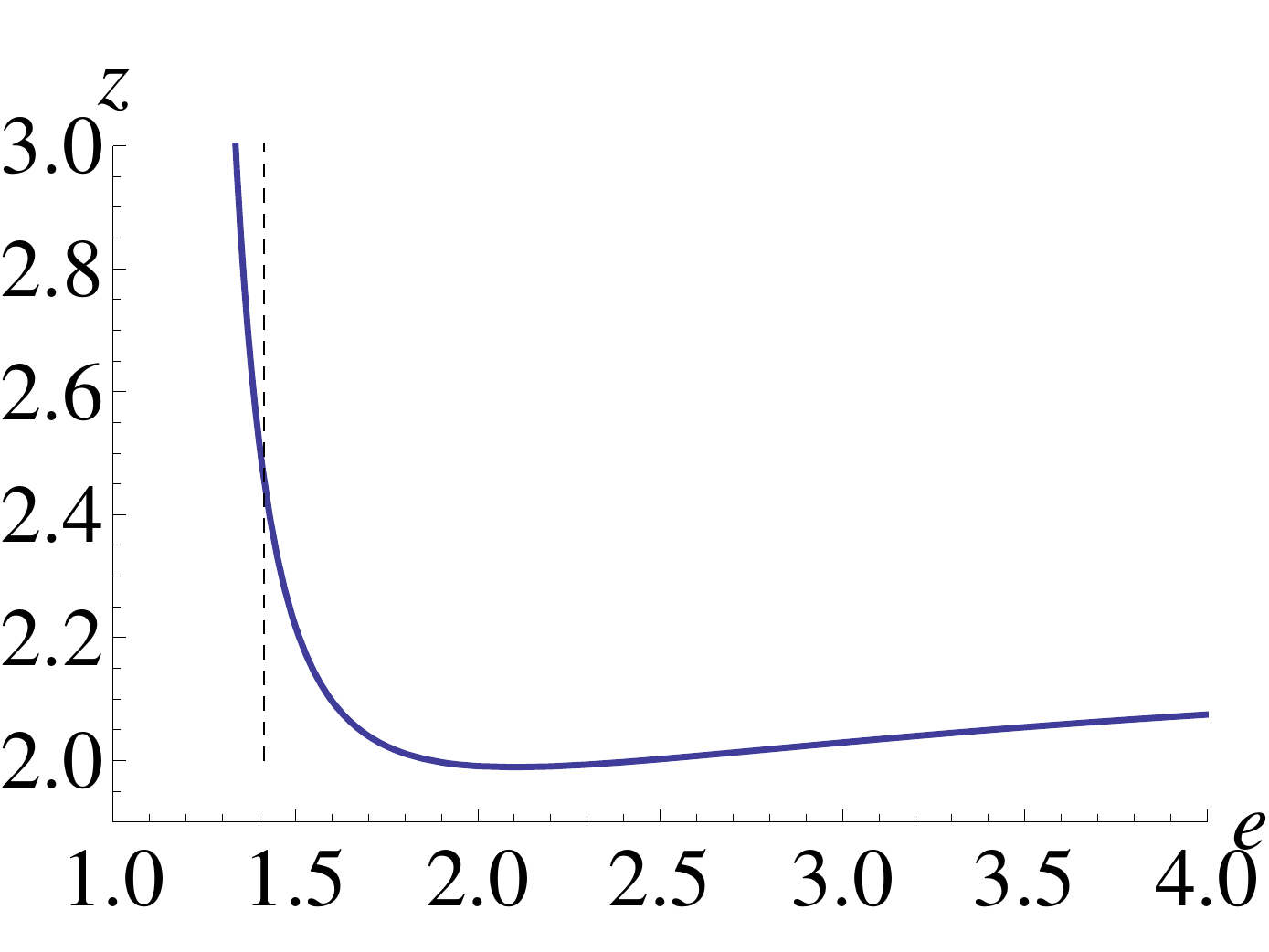}}
\caption{The scaling parameter $z$ for helical fixed point solutions with $k\ne 0$ of the form \eqref{helfp}, \eqref{helfpmet} as a function of $e$ with $m=2$ and
$e\gtrsim 1.13$.
The dashed vertical line is $e=\sqrt{2}$ and when $e>\sqrt{2}$ the AdS-RN black hole solution is unstable.}\label{figz}
\end{figure}

After substituting into the equations of motion we obtain a system of algebraic equations which can be solved numerically.
We find solutions provided $e\gtrsim 1.13$ and we have displayed the values of $z$ in figure \ref{figz}. In particular,
these fixed point solutions exist for all of the unstable AdS-RN black holes and hence can provide the IR limit of the 
zero temperature ground states of the $p$-wave black holes when $k\ne 0$. In fact this seems to be only true for the $p$-wave black holes with
$k>0$ at $T=0$.

\begin{figure}
\centering
{\includegraphics[width=6cm]{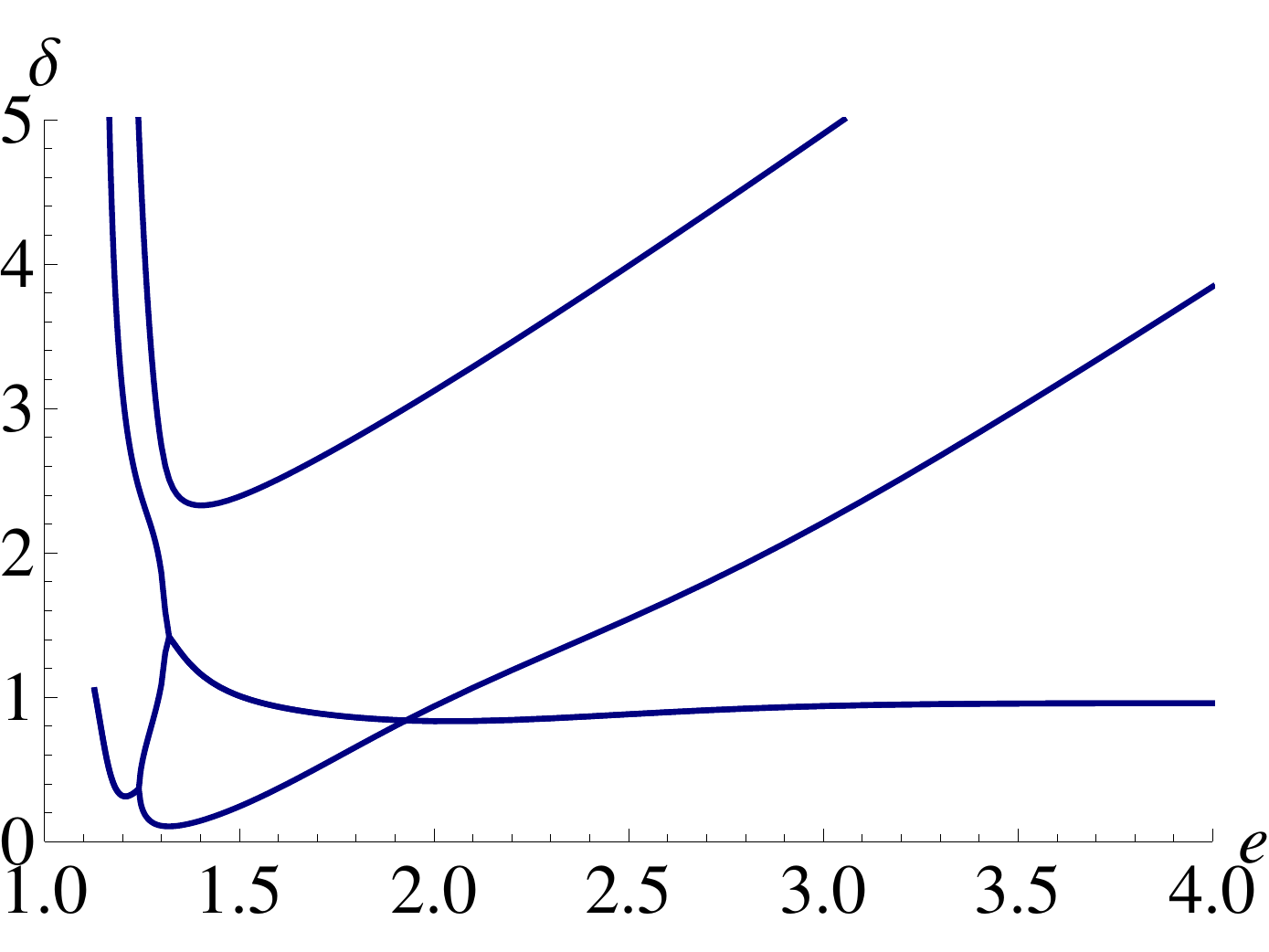}}
\caption{The scaling dimensions $\delta$ with positive real parts, in the perturbations about the Bianchi VII$_0$
fixed point solution
\eqref{helfp} with $k\ne 0$ as a function of $e$, with $m=2$. For 
$1.24\lesssim e\lesssim 1.31$ they are all real;  for other values of
$e$ two are real and two form a complex pair, related by conjugation. There is also one mode with $\delta=0$.}\label{figdel}
\end{figure}

To investigate the associated zero temperature domain wall solutions we need to examine perturbations about
the fixed point solution. We consider
\begin{align}
&g=r^{2}\,\left(L^{-2}+\lambda w_{1} r^{\delta} \right),\qquad\qquad f=\bar f_{0}\,r^{z-1}\left(1+\lambda w_{2}r^{\delta} \right)\,,\notag\\
&h=k\left(h_{0}+\lambda w_{3}r^{\delta}\right),\qquad\qquad
\alpha=\alpha_{0}+\lambda w_{4}r^{\delta}\,,\notag\\
&a=\bar f_{0}a_{0}r^{z}\,\left(1+\lambda w_{5}r^{\delta} \right),\,\,
\qquad c_{3}=kc_{3}^0r\,\left(1+\lambda w_{6}r^{\delta} \right).
\end{align}
After expanding the equations of motion at first order in $\lambda$ we obtain a homogeneous linear system of equations $\mathbf{E}\cdot\mathbf{w}=0$ \
where $\mathbf{E}$ is a $6\times 6$ matrix  that depends on $\delta$.
Demanding non-trivial solutions for $\mathbf{w}$ we determine the values of $\delta$ by solving the polynomial equation $\left| \mathbf{E}\right|=0$. The solutions come in five pairs with the sum of each pair equal to $-(2+z)$, as expected from a consideration of the scalar Laplacian
for the metric \eqref{helfpmet}. 
When $m=2$, $e=2$
the modes with non-negative real parts have 
$\delta_{1}=0$, $\delta_{2,3}\approx 0.83 \pm  1.01i$, $\delta_{4}\approx 0.94$ and $\delta_{5}\approx 3.12$.
In particular there is a pair related by complex conjugation\footnote{Such complex scaling dimensions are analogous to those appearing in \cite{Gubser:2009cg}.}. For different values of $e$ and $m=2$ this structure persists, with $\delta_{1}=0$ and one
complex-pair, except when $1.24\lesssim e\lesssim 1.31$ when all values of $\delta$ are real; see figure \ref{figdel}.

To construct domain wall solutions with \eqref{helfp} as the far IR behaviour, one needs to shoot out using
these five modes. Notice that
the mode with $\delta=0$ corresponds to the constant $\bar f_{0}$. This leads to five (real) parameters in the IR. With the eight parameters
mentioned for the UV discussed in \eqref{uvexpp} and two scaling symmetries from \eqref{eq:symmetries1}, 
we deduce that the domain wall solutions will be specified by a single parameter
which we can take to be the wave-number $k$. Note that such domain walls were constructed in 
\cite{Donos:2012gg} for $m=1.7$ and $e=1.88$ (a case where all $\delta$ are real). It would be interesting to determine whether
or not these domain wall solutions exist for all values of $k\ne 0$; as we will discuss later, it is possible that they only exist for $k>0$.

\subsubsection{Scaling solutions with $k= 0$}\label{scalkzp}
When $k=0$ the equations of motion admit scaling solutions of the form 
\begin{align}\label{helfpkz}
g=L^{-2}r^2,\quad f=\bar f_0r^{z-1},\quad h=\alpha_0^{-1}r^{1-\gamma},\quad e^\alpha=\alpha_0r^\gamma\quad
a=\bar f_0a_0 r^z,\quad c_3= c_3^0\alpha_0^{-2}r^{2(1-\gamma)}\,.
\end{align}
where $z, L,\bar f_0, \gamma,\alpha_0,a_0$ and $c_3^0$ are constants
and we note that we are within the consistently truncated ansatz with $h=re^{-\alpha}$.
Explicitly, the full solution reads
\begin{align}\label{helfpkzex}
ds^{2}&=-(\bar f_0^2L^{-2})r^{2z}dt^{2}+L^2\frac{dr^{2}}{r^2}+\alpha_0^{-2}r^{2(1-\gamma)}\left(dx_{1}^{2}+dx_3^2\right)+\alpha_0^2r^{2(1+\gamma)}dx_2^2\,,\notag\\
C&=(-i\tfrac{2(1-\gamma)\bar f_0 \alpha_0 c_3^0}{mL^2})r^{z+1+\gamma}dt\wedge dx_2
+(\tfrac{ea_0 \alpha_0c_3^0L^2}{m })r^\gamma dr\wedge dx_{2}+(c_3^0\alpha_0^{-2})r^{2(1-\gamma)}\,dx_{1}\wedge dx_{3}\,,\nn
A&=(\bar f_0a_0)r^zdt\,,
\end{align}
By scaling $t$ we can set $\bar f_0=1$ and by suitably scaling $x_1,x_3$ and $x_2$ we can set $\alpha_0=1$ too. 
These fixed point solutions are
invariant under the anisotropic scaling 
\begin{align}
r\to \lambda^{-1}r,\quad t\to\lambda^z t,\quad x_{1,3}\to \lambda^{1-\gamma} x_{1,3},\quad x_2\to \lambda^{1+\gamma}x_2\,.
\end{align}
They are also translationally invariant in all three spatial directions and rotationally invariant in the $x_1,x_3$ plane; they 
are somewhat similar to some solutions constructed in \cite{Taylor:2008tg}.

\begin{figure}
\centering
{\includegraphics[width=6cm]{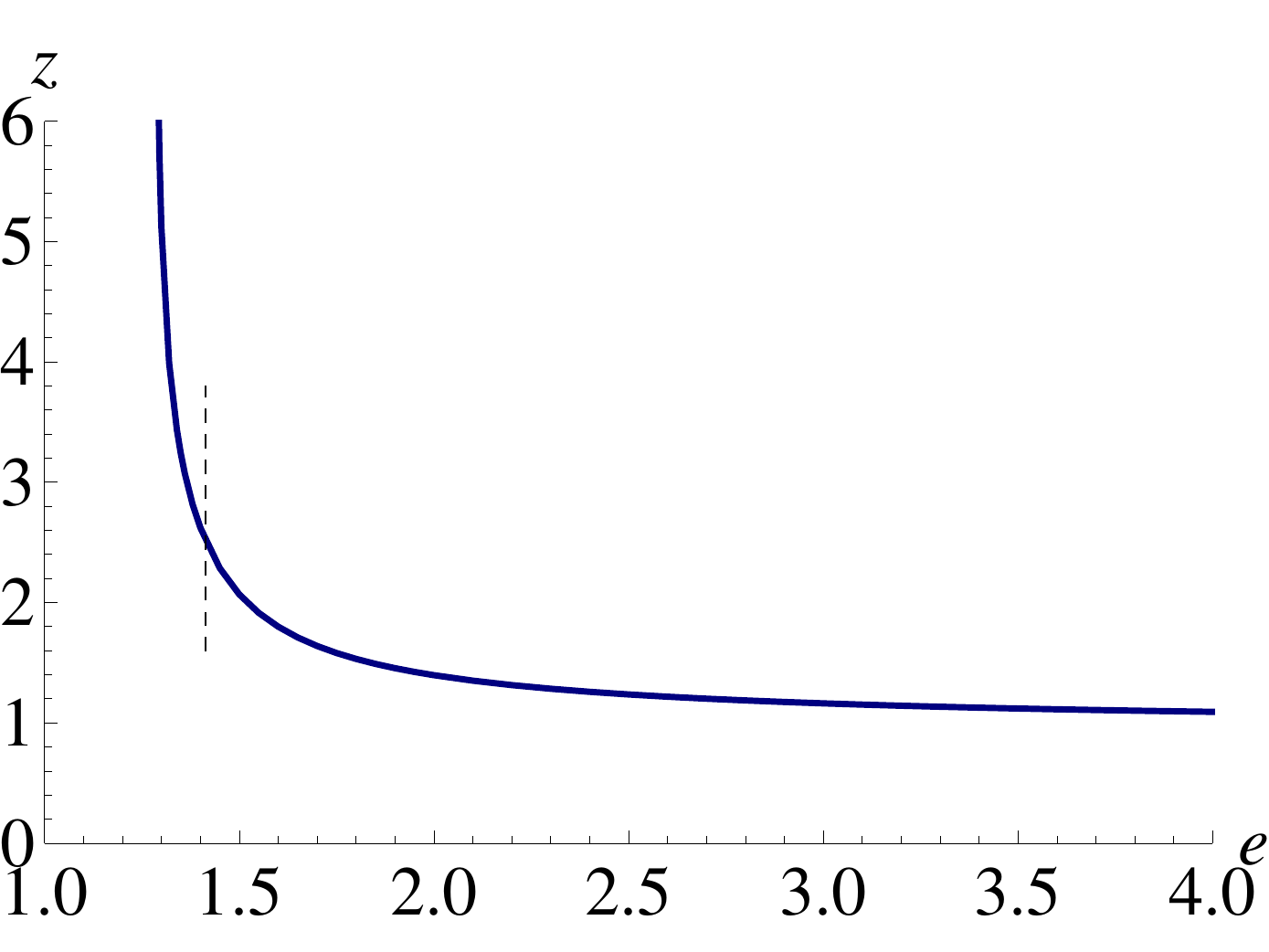}}\qquad\qquad
{\includegraphics[width=6cm]{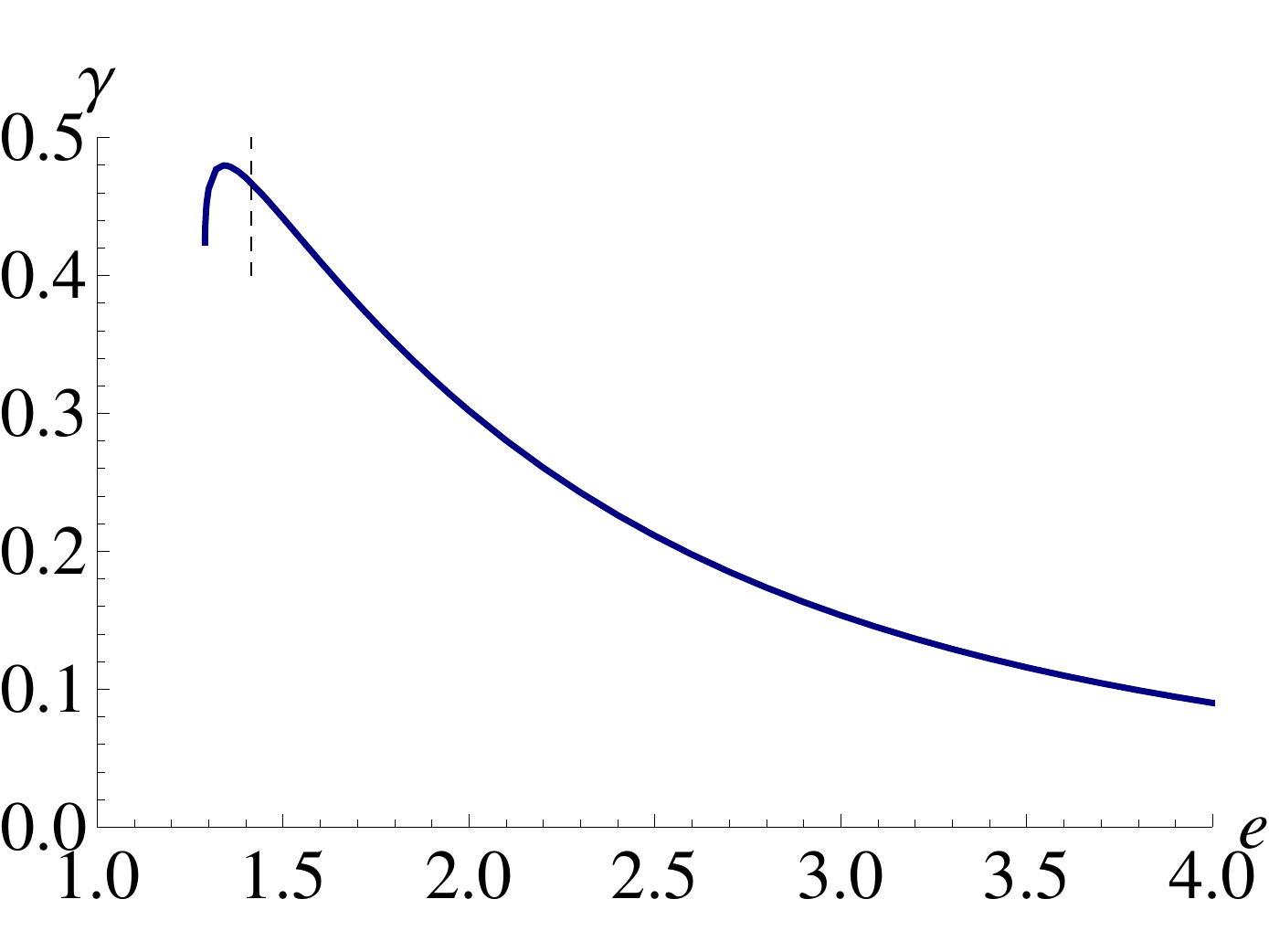}}
\caption{The scaling parameters $z,\gamma$ for helical fixed point solutions with $k=0$ of the form \eqref{helfpkzex} as a function of $e$ with $m=2$ and
$e\gtrsim 1.29$.
The dashed vertical line is $e=\sqrt{2}$ and when $e>\sqrt{2}$ the AdS-RN black hole solution is unstable.}\label{figz2}
\end{figure}

After substituting into the equations of motion we obtain a system of algebraic equations which can be solved numerically.
We find solutions provided $e\gtrsim 1.29$ and we have displayed the values of $z$ and $\gamma$ in figure \ref{figz2}. In particular we notice that
these fixed point solutions exist for all of the unstable AdS-RN black holes and hence can provide the IR limit of the 
zero temperature ground states when $k=0$.

\begin{figure}
\centering
{\includegraphics[width=6cm]{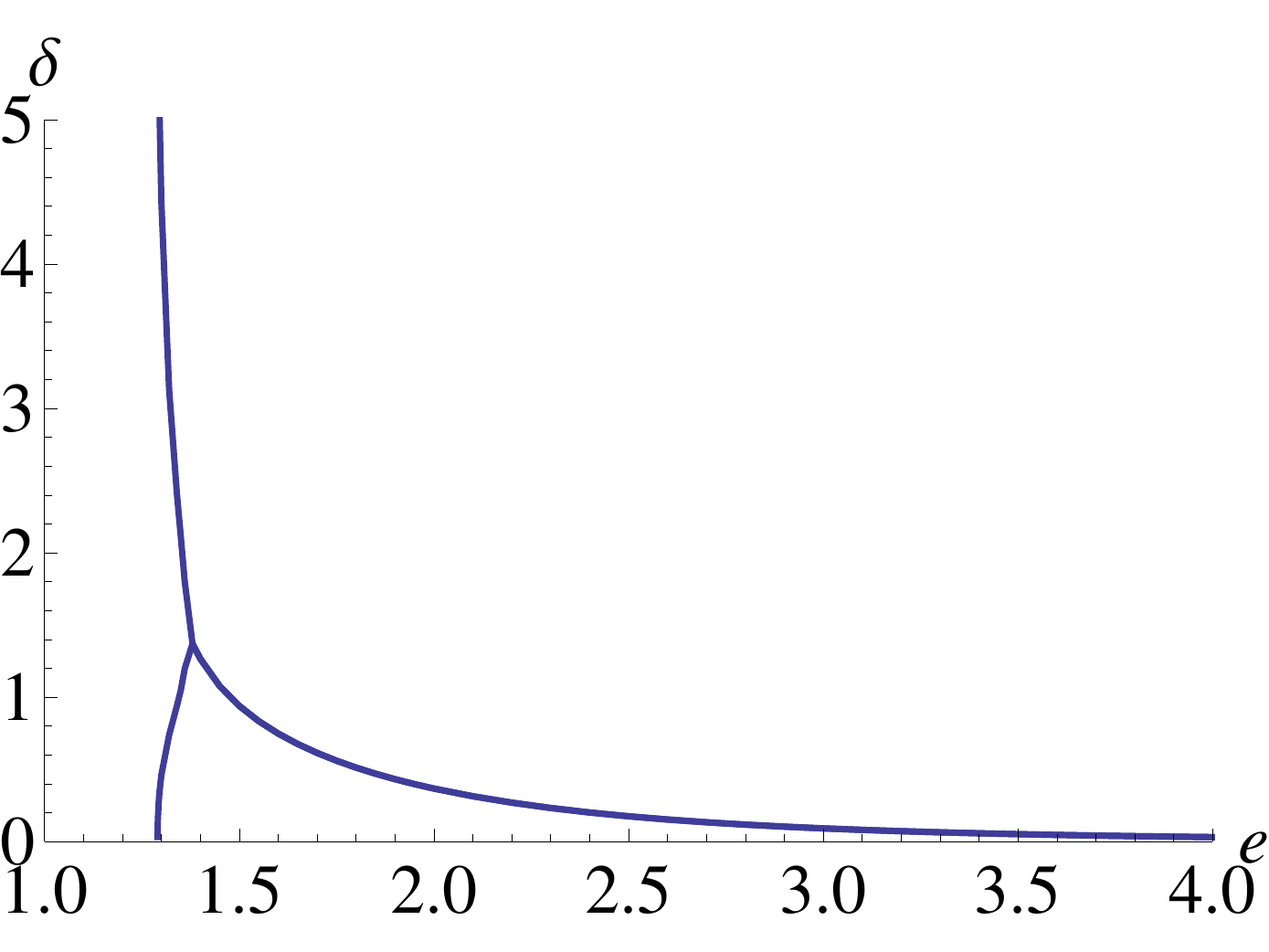}}
\caption{The scaling dimensions $\delta$ with positive real parts, in the perturbations about the helical fixed point solution
\eqref{helfpkzex} with $k=0$ as a function of $e$, with $m=2$. For $e\lesssim 1.36$ they are both real and for $e\gtrsim 1.36$, they coalesce to
form a complex conjugate pair. There are also two modes with $\delta=0$.}\label{figdel2}
\end{figure}

To investigate zero temperature domain wall solutions we need to examine perturbations about
the fixed point solution. We consider
\begin{align}
&g=r^{2}\,\left(L^{-2}+\lambda w_{1} r^{\delta} \right),\qquad\qquad\qquad f=\bar f_{0}\,r^{z-1}\left(1+\lambda w_{2}r^{\delta} \right)\,,\notag\\
&h=re^{-\alpha},\qquad\qquad\qquad\qquad\qquad
e^\alpha=\alpha_0r^\gamma(1+\lambda w_{3}r^{\delta})\,,\notag\\
&a=\bar f_{0}a_{0}r^{z}\,\left(1+\lambda w_{4}r^{\delta} \right),\,\,
\qquad\qquad c_{3}=c_{3}^0\alpha_0^{-2}r^{2(1-\gamma)}\,\left(1+\lambda w_{5}r^{\delta} \right).
\end{align}
After expanding the equations of motion at first order in $\lambda$ we obtain a homogeneous linear system of equations $\mathbf{E}\cdot\mathbf{w}=0$ \
where $\mathbf{E}$ is a $5\times 5$ matrix  that depends on $\delta$.
Demanding non-trivial solutions for $\mathbf{w}$ we determine the values of $\delta$ by solving the polynomial equation $\left| \mathbf{E}\right|=0$. The solutions come in four pairs with the sum of each element of a pair equal to $-(3+z-\gamma)$, as expected. 
When $m=2$, $e=2$
the modes with non-negative real parts have 
$\delta_{1,2}=0$,and $\delta_{3,4}\approx 0.37 \pm 0.61 i$.
In particular there is a pair related by complex conjugation. 
The two modes with $\delta=0$ are associated with the parameters $\bar f_0$ and $\alpha_0$ in \eqref{helfpkzex}.
For different values of $e$ and $m=2$ this structure persists, with $\delta_{1,2}=0$ and one
complex-conjugate pair, except when $e\lesssim 1.36$ when all values of $\delta$ are real; see figure \ref{figdel2}.

To construct domain wall solutions with \eqref{helfp} as the far IR behaviour, one needs to shoot out using
these four modes in the IR. With the six parameters
mentioned for the UV discussed in \eqref{uvexpp} (recall that for $k=$ we have $c_\alpha=c_h=0$)
and two scaling symmetries from \eqref{eq:symmetries1}, 
we can deduce that the domain wall solutions, if they exist, will be unique. 

\subsubsection{Domain walls with $AdS_{5}$ in the IR}\label{pads5ir}
We now consider the possibility of constructing domain walls that interpolate between $AdS_5$ in the UV and the {\it same} $AdS_5$ in the IR.
Since one usually shoots away from the UV and IR using relevant and irrelevant modes, respectively,
and since we have the same $AdS_5$ in the UV and the IR, one's first thought is that this won't be possible.
On the other hand a construction was made in \cite{Horowitz:2009ij}
using a massless complex scalar. In that work, the scalar was a constant in the IR and this
provided mass for the gauge-field, transforming it into an irrelevant operator
and enabling the construction of superconducting domain walls interpolating between the same $AdS$ spacetime.
Another construction in the context of $p$-wave superconductors with bulk $SU(2)$ gauge fields appears in \cite{Basu:2009vv}.

Here we will describe a new construction that instead exploits the $k$ dependence of the relevant modes. 
Specifically, we have found that one can
build up an expansion about the $AdS_5$ spacetime as follows. We consider, as $r\to 0$,
\begin{align}
&g=r^2+\delta g,\qquad f=\bar f_0+\delta f,\qquad h=h_0 r+\delta h,\nn
&\alpha=\delta \alpha,\qquad c_3=\delta c_3,\qquad a=a_0+\delta a,
\end{align}
where $\bar f_0, h_0$ and $a_0$ are constants, associated with marginal deformations about the $AdS_5$. 
After substituting into the equations of motion, at first order in
the perturbation we find that we need to solve second order linear ODEs for $\delta\alpha$ and $\delta c_3$. Furthermore, we find that
the following solution can be developed:
\begin{align}
\delta\alpha=\alpha_{0}\frac{e^{-2{k}/h_0r}}{r^{3/2}}(1+{\cal O}(r)),\qquad
\delta c_3=c_3^0\frac{e^{-{kx}/h_0r}}{r^{1/2}}(1+{\cal O}(r))\,,
\end{align}
where $x=(1-a_0^2h_0^2e^2/\bar f_0^2k^2)^{1/2}$ and $\alpha_0$, $c_3^0$ are constants. 
Indeed going to second order in the perturbation (i.e. keeping terms
with $(c_3^0)^2$, $c_3^0\alpha_{0}$ and $\alpha_{0}{^2}$) we find
\begin{align}\label{ads5exp}
&g=r^{2}-\frac{(c_3^{0})^{2}}{6h_{0}{k}}x\frac{e^{-2{kx}/h_0r}}{r^2}(1+{\cal O}(r))+\alpha_0^2\frac{e^{-4k/h_0r}}{r}(1+{\cal O}(r))
\cdots\,,\notag\\
&f=\bar f_{0}+\frac{(c_3^0)^2\bar f_0x}{12h_0 k }\frac{e^{-2{kx}/h_0r}}{r^4}(1+{\cal O}(r))
-\frac{\alpha_0^2\bar f_0}{2}\frac{e^{-4k/h_0r}}{r^3}(1+{\cal O}(r))
\cdots,\nn
&h=h_{0}r
-\frac{(c_3^{0})^{2}a_{0}^{2}h_0^3e^{2}}{8\bar f_{0}^{2}k^4 x^2}\frac{e^{-2{kx}/h_0r}}{r^2}(1+{\cal O}(r))
-\frac{\alpha_0^2 h_0}{2}\frac{e^{-4k/h_0r}}{r^2}(1+{\cal O}(r))
+\dots,\nn
&\alpha=\alpha_{0}\frac{e^{-2{k}/h_0r}}{r^{3/2}}(1+{\cal O}(r))
-\frac{(c_3^{0})^{2}}{8k^2}\frac{e^{-2{kx}/h_0r}}{r^3}(1+{\cal O}(r))+\cdots\,,\nn
&c_{3}=c_3^0\frac{e^{-{kx}/h_0r}}{r^{1/2}}(1+{\cal O}(r))
+\alpha_0c_3^0\frac{e^{-{kx}/h_0r}e^{-2k/h_0r}}{r^{2}}(1+{\cal O}(r))
\cdots,\nn
&a=a_{0}-\frac{(c_3^{0})^{2}\bar f_{0}e}{4m{k}^{2}x}\frac{e^{-2{kx}/h_0r}}{r^2}(1+{\cal O}(r))+\cdots\,.
\end{align}
Higher order corrections will involve higher powers of the exponential terms.
The key aspect of this expansion is that the $k$ dependent exponentials allow the corresponding terms to remain small as
$r\to 0$. Notice that the $k=0$ limit of this expansion is not well-defined and this is consistent with the fact that it would
not be possible to develop a $k$ independent expansion in the IR associated with $\delta\alpha$ and $\delta c_3$ which are dual
to relevant operators.  
Note also, that in contrast to \cite{Horowitz:2009ij}, the function $c_3$ governing the superconductivity is going to zero as $r\to 0$.
Finally, we comment that a related expansion was, independently, recently used in a different context in \cite{Chesler:2013qla}.

Observe that this expansion has five IR parameters and so we expect that when combined with the eight UV parameters 
in \eqref{uvexpp} and the two scaling symmetries from \eqref{eq:symmetries1}, there will be a one-parameter family of domain wall solutions, which can be labelled by $k$.

\subsection{The $T\to 0$ limit of the $p$-wave black holes}
\label{tzerolimp}

For a given temperature $T\ne 0$ the properties of the $p$-wave black holes depend continuously on $k$. For example,
one can see this in the behaviour of the free-energy plot in figure \ref{figtwo}(d).
As $T\to 0$, all of the $p$-wave black holes appear to approach zero entropy ground states, $s\to 0$.
However, the precise domain wall solutions that the black holes approach at $T=0$ depends
on the sign of $k$, as we shall discuss. We note at the outset that in many cases the $T\to 0$ behaviour 
only manifests itself at extremely low temperatures, which becomes very challenging to analyse numerically.
The main strategy is to seek evidence that the entropy density scales with temperature consistent with the scaling
expected from a domain wall solution that hits a scaling solution in the far IR.

\subsubsection{$k>0$}
For $m=1.7$ and $e=1.88$ it was shown in \cite{Donos:2012gg} that at $T=0$ 
the black holes with $k>0$ approach domain wall solutions which
in the far IR approach scaling solutions with a Bianchi VII$_0$ symmetry \eqref{helfp}. 
More precisely this was shown for the preferred branch of black holes, which had $k> 0$ at $T=0$, and 
it was also shown for black holes with slightly smaller and also larger values of $k$. 
In particular, explicit domain wall solutions were constructed in detail for these values of $k$
but smaller values of $k$, including $k\le 0$, were not studied.

For $m=2$ we again seem to find that for $k>0$ the black holes approach similar domain wall solutions
with the Bianchi VII$_0$ fixed points \eqref{helfp} in the IR. Amongst other things, we have checked for 
the thermodynamically preferred branch black holes with $e=2$, 
which have $k=0.30$ at $T=0$, 
that the temperature dependence of the entropy has the appropriate scaling behaviour. Explicitly,
if the black hole solutions approach the fixed points \eqref{helfp} in the IR, then a simple argument based on dimensional analysis
(e.g. see \cite{Hartnoll:2009ns}) shows that the entropy density should obey the scaling
\begin{align}\label{entrelone}
s\propto T^{2/z}\,,
\end{align}
where the factor of two arises because only two of the spatial directions are involved in the scaling \eqref{knzsc}.
For $e=2$ the appropriate value of $z$ is given by $z\sim 1.99$ (see figure \ref{figz}).
We have plotted our results in figure \ref{fignine}(a), and we note that we went down to temperatures of the order $3\times 10^{-8}$.

\subsubsection{$k=0$}

\begin{figure}
\centering
\subfloat[]{\includegraphics[width=5cm]{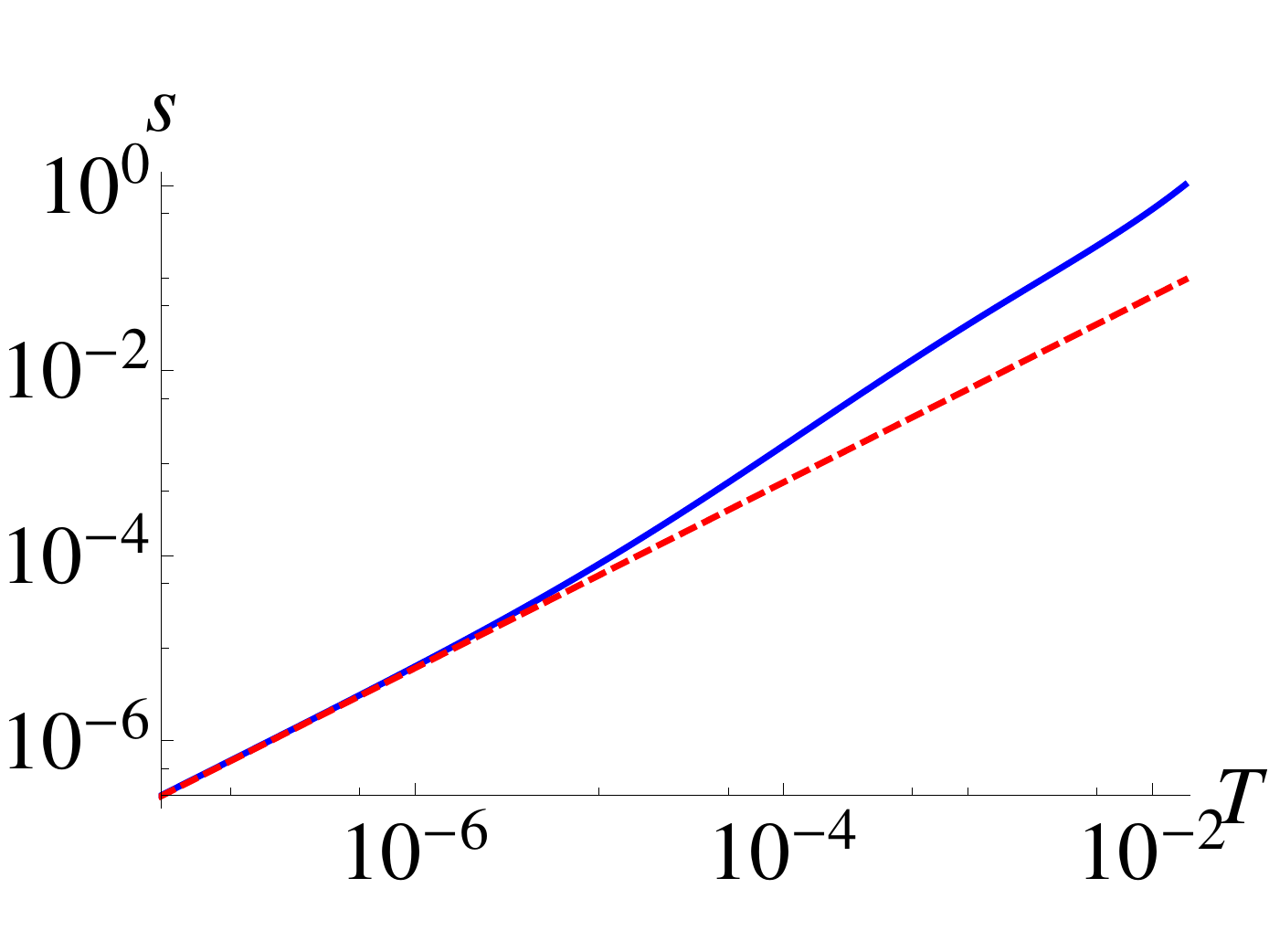}}\hskip2em
\subfloat[]{\includegraphics[width=5cm]{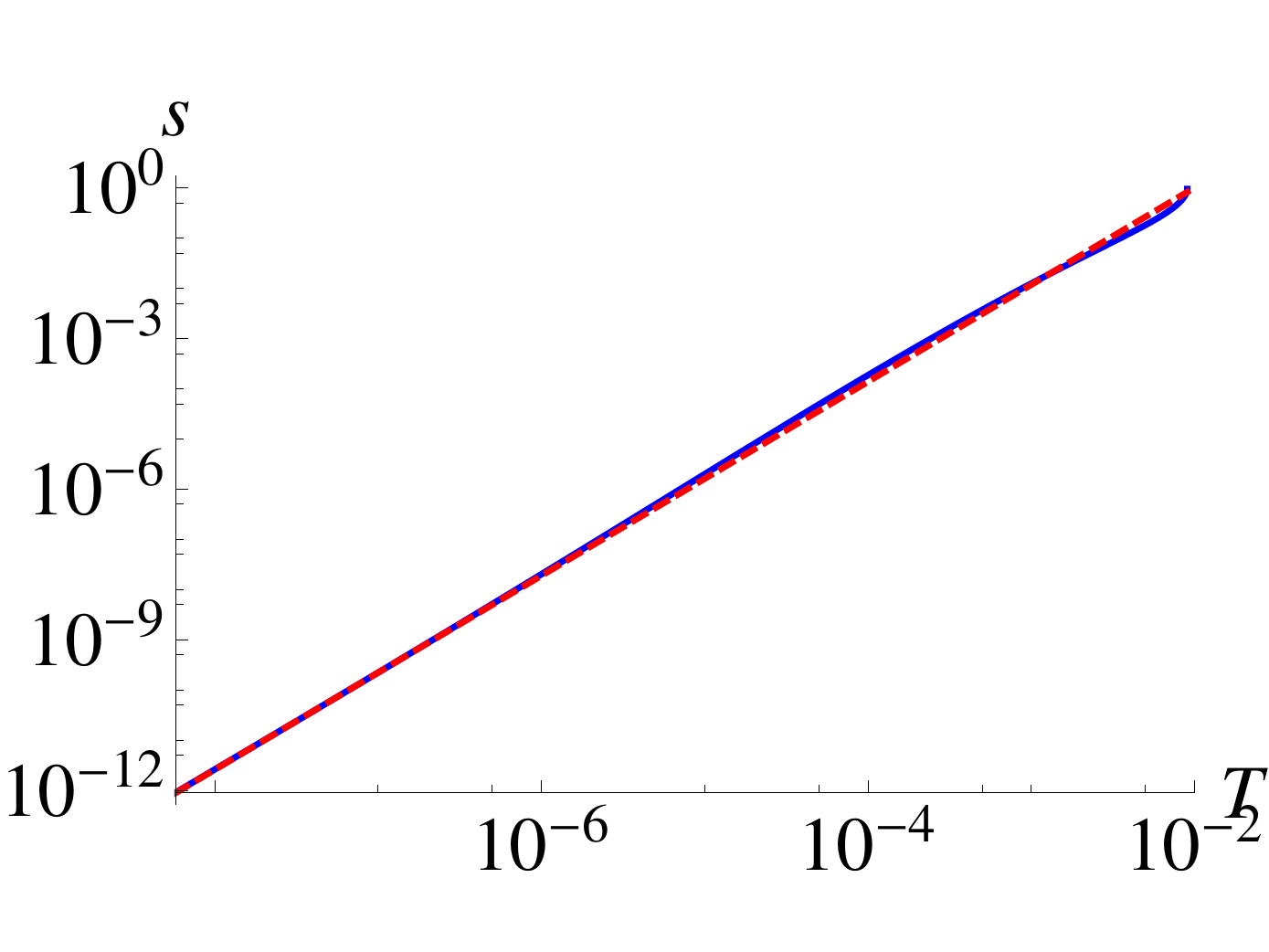}}
\caption{The low temperature behaviour of the entropy density $s$ for the 
$p$-wave black holes for $m=2$ and $e=2$. 
The blue line in panel (a) is for the thermodynamically preferred branch (the red line in figure \ref{figtwo}a). It is consistent
with the solutions approaching $k>0$ domain walls with Bianchi VII$_0$ fixed points 
in the IR \eqref{helfp} with $s\propto T^{2/z}$ and $z=1.99$ (marked by the red dashed line).
The blue line in panel (b) is for the {\it non-preferred} $k=0$ branch, consistent with 
the solutions approaching the scaling behaviour $s\propto T^{(3-\gamma)/z}$ and $z\sim1.40$, $\gamma\sim0.0302$
associated with the scaling solution \eqref{helfpkzex} (marked by the red dashed line). 
We have set $\mu=1$.}\label{fignine}
\end{figure}

If the $T\to 0$ black hole solutions approach $k=0$ domain walls with fixed points \eqref{helfpkzex} in the IR, 
then dimensional analysis implies that the entropy density should behave like 
\begin{align}\label{entreltwo}
s\propto T^{(3-\gamma)/z}\,.
\end{align}
For $m=2$ and $e=2$ we have found that the $k=0$ branch of black holes (which is not thermodynamically preferred) does in fact approach
such scaling behaviour, as illustrated in figure \ref{fignine}(b). For this case the scaling behaviour again starts to manifest itself at very low temperatures.

\subsubsection{$k<0$}
The $T\to 0$ limits of the $p$-wave black holes with $k<0$ are the most difficult to analyse numerically. 
Nevertheless, our analysis of the behaviour of the
functions near the horizon indicate that the black holes approach the $AdS_5$ vacuum in the far IR, but the temperatures are extremely low. Indeed we have constructed black holes with temperatures  of the order $10^{-8}$ where some evidence of
an $AdS_5$ region appears to be building up near the black hole horizon. However, to get a conclusive picture one needs to go to even lower temperatures and this will be left for future work.
The fact that we have provided evidence that there are appropriate domain wall solutions, using the expansion \eqref{ads5exp}, lends support to
the conclusion that an $AdS_5$ appears in the far IR at $T=0$.

\subsubsection{Summary}
Our results indicate the following picture for the $T\to 0$ limits of the $p$-wave black holes.
For $k>0$ they approach domain walls with Bianchi VII$_0$ scaling in the IR. For $k=0$ they approach domain walls with anisotropic scaling
\eqref{helfpkzex} in the IR. For $k<0$ they approach $AdS_5$ in the IR. 

It seems reasonable to conjecture that the domain
walls with Bianchi VII$_0$ fixed points in the IR in fact only exist for $k>0$ and similarly the domain walls with $AdS_5$ in the IR only exist
for $k<0$. It will be interesting to investigate this further.

\subsubsection{$\partial_T k=0$ as $T\to 0$}\label{littlearg}
A feature of the thermodynamically preferred red branches in figure \ref{figtwo}, is that $\partial_T k=0$ as $T\to 0$.
To conclude this section we explain this fact using the following simple argument.
The free energy density $w$ is a function of $T,k$ and $\mu$. However, we are interested in holding $\mu$ fixed and indeed we
have set $\mu=1$ in our plots. The thermodynamically preferred black hole solutions satisfy $\partial_k w=0$ at fixed $T$
(giving the condition $c_h=0$) and this specifies $k=k(T)$ along this branch. Thus, along this branch we compute
\begin{align}
0=\frac{d}{dT}(\partial_k w)&=\partial_{Tk}^2w+\partial_{k}^2 w\partial_Tk\,,\nn 
&=-\partial_k s+\partial_{k}^2 w\partial_Tk\,.
\end{align}
Since the thermodynamically preferred branch minimises the free energy we have (generically) $\partial_{k}^2 w>0$.
As $T\to 0$ if the preferred solutions have $\partial_k s=0$, for example if $s\to 0$ in some open subset as we have for the $p$-wave black hole solutions, 
then we can conclude that $\partial_Tk =0$ as $T\to 0$.

Clearly this argument applies more generally and indeed the behaviour $\partial_T k=0$ as $T\to 0$ can be observed for the preferred branch of striped black holes
constructed in \cite{Rozali:2012es,Donos:2013wia,Withers:2013loa,Withers:2013kva,Rozali:2013ama} (see figure 4 of \cite{Withers:2013kva}). We also note that the argument
can be easily extended to black holes that are spatially modulated in more than one direction.

\section{$(p+ip)$-wave black holes}\label{sectionp+ip}
The ansatz we shall consider for the black holes describing holographic $(p+ip)$-wave superconductors
is given by 
\begin{align}
\label{eq:ansatz}
ds^{2}&=-g f^2 dt^2+g^{-1}{dr^2}+h^2(dx_1+Q dt)^2+r^2(dx_2^2+dx_3^2)\,,\nonumber\\
A&=a dt+b dx_1\,,\nonumber\\
C&=e^{-i k x_1}(i c_1dt+c_2dr+ic_3 dx_1) \wedge (dx_2-i dx_3)\,,
\end{align}
where 
$f,g,h,Q,a,b$ and $c_i$ are nine functions of the radial coordinate $r$ and $k$ is a constant corresponding to
the wave-number of the $(p+ip)$-wave order. The ansatz is invariant under time translations and
when $Q\neq 0$ the spacetime is stationary but not static. The ansatz is invariant under translations
in the $x_2, x_3$ directions and also in the $x_1$ direction when combined with a constant gauge transformation:
$x_1\to x_1+c$, $C\to e^{ikc}C$. The ansatz is invariant under rotations in the $(x_2,x_3)$-plane when combined with a
gauge transformation or, when $k\ne 0$, a translation in the $x_1$ direction. Observe that a gauge transformation can be used to 
eliminate the phase $e^{ik x_1}$ appearing in $C$ by making the shift $b\to b-k/e$. We will return to this point later. 

The equations of motion we are interested in are obtained by substituting
the ansatz \eqref{eq:ansatz} into \eqref{fulleom}. They can also be 
obtained by substituting the ansatz  \eqref{eq:ansatz} directly into the action \eqref{eq:action} to obtain
\begin{align}
\label{eq:action1}
S&=\int d^5x r^2hf\Bigg\{-g''-g'\left(\frac{3f'}{f}  +\frac{2 h'}{h} +\frac{4}{r} \right)+12+\frac{ h^2 Q'^2}{2 f^2}\nn
&-2g\left[ \frac{f''}{f}+\frac{f'}{f}\left(\frac{2}{r}+\frac{h'}{h}\right)+\frac{h''}{h}+\frac{2h'}{rh}+\frac{1}{r^2}  \right]
+\frac{a'^2}{2 f^2}-\frac{1}{2}\left(\frac{g}{h^2}-\frac{Q^2}{f^2}\right)b'^2-\frac{Qa'b'}{f^2}\nn
&+\frac{c_1^2}{r^2f^2g}-\frac{gc_2^2}{r^2}+\frac{1}{r^2}
\left(\frac{Q^2}{f^2 g}-\frac{1}{h^2}\right)c_3^2-\frac{2Qc_1c_3}{r^2f^2g}\Bigg\}\nn
  &+\frac{1}{m}\int d^5x\Bigg\{{c_1 c_ 3'}-{c_3 c_1'}+{2 e (a c_2 c_3-bc_1c_2)}+{2 k c_1
  c_ 2}\Bigg\}\,,
\end{align}
and then varying with respect to the nine functions, holding $k$ fixed. 
We find that $f$ and $g$ satisfy first order differential equations and that $h,Q,b,a$ and $c_3$ satisfy second order
equations with $c_1$ and $c_2$ completely determined in terms of $c_3$ via
\begin{align}
\label{eq:c1c2}
c_1&=\frac{[ea(eb-k)+m^2 h^2 Q]c_3-mhfgc_3'}{m^2 h^2+(eb-k)^2}\,,\nonumber\\
c_2&=\frac{[ea-Q(eb-k)]mhc_3 +fg(eb-k)c_3'}{fg[m^2 h^2+(eb-k)^2]}\,.
\end{align} 
We observe that our ansatz, and hence the equations of motion, are left invariant under the following three scaling symmetries:
\begin{align}
\label{eq:symmetries}
&r \to \lambda r\,, \quad (t,x_2,x_3) \to \lambda^{-1}(t,x_2,x_3)\,, \quad g\to \lambda^2 g\,, \quad Q \to \lambda Q\,, \quad a \to \lambda a\, ,\quad c_3 \to \lambda c_3\,; \nonumber\\
&x_1 \to \lambda^{-1} x_1\,, \quad h \to \lambda h\,, \quad Q\to \lambda^{-1} Q\,, \quad  k \to \lambda k\,, \quad b \to \lambda b\, , \quad c_3 \to \lambda c_3\,;\nonumber\\
&t \to \lambda t\,, \quad f \to \lambda^{-1}f\,, \quad a\to \lambda^{-1} a\,, \quad Q \to \lambda^{-1} Q\,; \end{align}
where $\lambda$ is a constant.

\subsection{Asymptotic and near-horizon expansions}
We will be interested in black hole solutions that asymptotically approach $AdS_5$ in the UV and are dual to $d=4$ phases where the symmetry breaking is spontaneously generated. By analysing the equations of motion we can construct the following asymptotic expansion as $r \to \infty$:
\begin{align}
\label{eq:UVexp}
g&=r^2 (1-{M}{r^{-4}}+\cdots)\,,\qquad
f=f_0 (1-{c_h}{r^{-4}}+\cdots)\,,\nonumber\\
h&=r (1+{c_h}{r^{-4}}+\cdots)\,,\qquad
Q={c_Q}{r^{-4}}+\cdots\,,\nonumber\\
a&=f_0 (\mu+{q}{r^{-2}}+\cdots)\,,\qquad
b={c_b}{r^{-2}}+\cdots\,,\nonumber\\
c_3&={c_{c_3}}{r^{-|m|}}+\cdots\,.
\end{align} 
At a convenient juncture we will use the symmetries \eqref{eq:symmetries} to set $f_0=\mu=1$. The UV data is then specified by seven parameters $M,c_h,c_Q,q,c_b,c_{c_3}$ and $k$. Note that we have fixed the asymptotic fall-off 
of $h$ in \eqref{eq:UVexp} so we can no longer use \eqref{eq:symmetries} to scale $k$. 
The fall-off of $c_{c_3}$ and $b$ is appropriate for the spontaneous appearance of the $(p+ip)$-wave order, labelled by wave-number $k$.
In particular, the fall-off of $b$ corresponds to the absence of a source for the global $U(1)$ symmetry in the boundary field theory. 
Observe, however, that if we implement the gauge transformation mentioned above to eliminate the phase $e^{-ik x_1}$ appearing in the 
two-form $C$,  then we should instead impose the UV boundary condition $b=-k/e +{c_b}r^{-2}$.
The holographic interpretation of the other parameters appearing in the ansatz will become clear when we derive the holographic stress tensor and
current, below.

At the black hole horizon, located at $r=r_+$, the functions have the analytic expansion
\begin{align}
\label{eq:IRexp}
g&=g_+(r-r_+)+\cdots\,,\qquad
f=f_++\cdots\,,\nonumber\\
h&=h_++\cdots\,,\qquad\qquad\quad
Q=Q_+(r-r_+)+\cdots\,,\nonumber\\
a&=a_+(r-r_+)+\cdots\,,\qquad
b=b_++\cdots\,,\nonumber\\
c_3&=c_{3+}+\cdots\,.
\end{align} 
Regularity of the metric at the black hole horizon can easily be seen by using the in-going Eddington-Finkelstein coordinates $v,r$ where $v\approx t+(g_+ f_+)^{-1} ln(r-r_+)$. The full IR expansion is fixed in terms of the seven constants $f_+, h_+,Q_+, a_+,b_+,c_{3+}$ and $r_+$. In particular, the coefficient $g_+$ is fixed by these constants:
\begin{equation}
g_+=r_+(4-\frac{a_+^2}{6 f_+^2})-\frac{c_{3+}^2}{6 r_+ h_+^2}\,.
\end{equation}

After fixing the symmetries \eqref{eq:symmetries}, we have seven UV parameters and seven IR parameters. We have two first order differential equations and five which are second order, so a solution is fixed by twelve integration constants. Thus, generically, we expect a two-parameter family black hole solutions, which we will label  by the wave-number, $k$, and temperature, $T$.

\subsection{Numerical solutions}
\begin{figure}
\centering
\subfloat[]{\includegraphics[width=4.5cm]{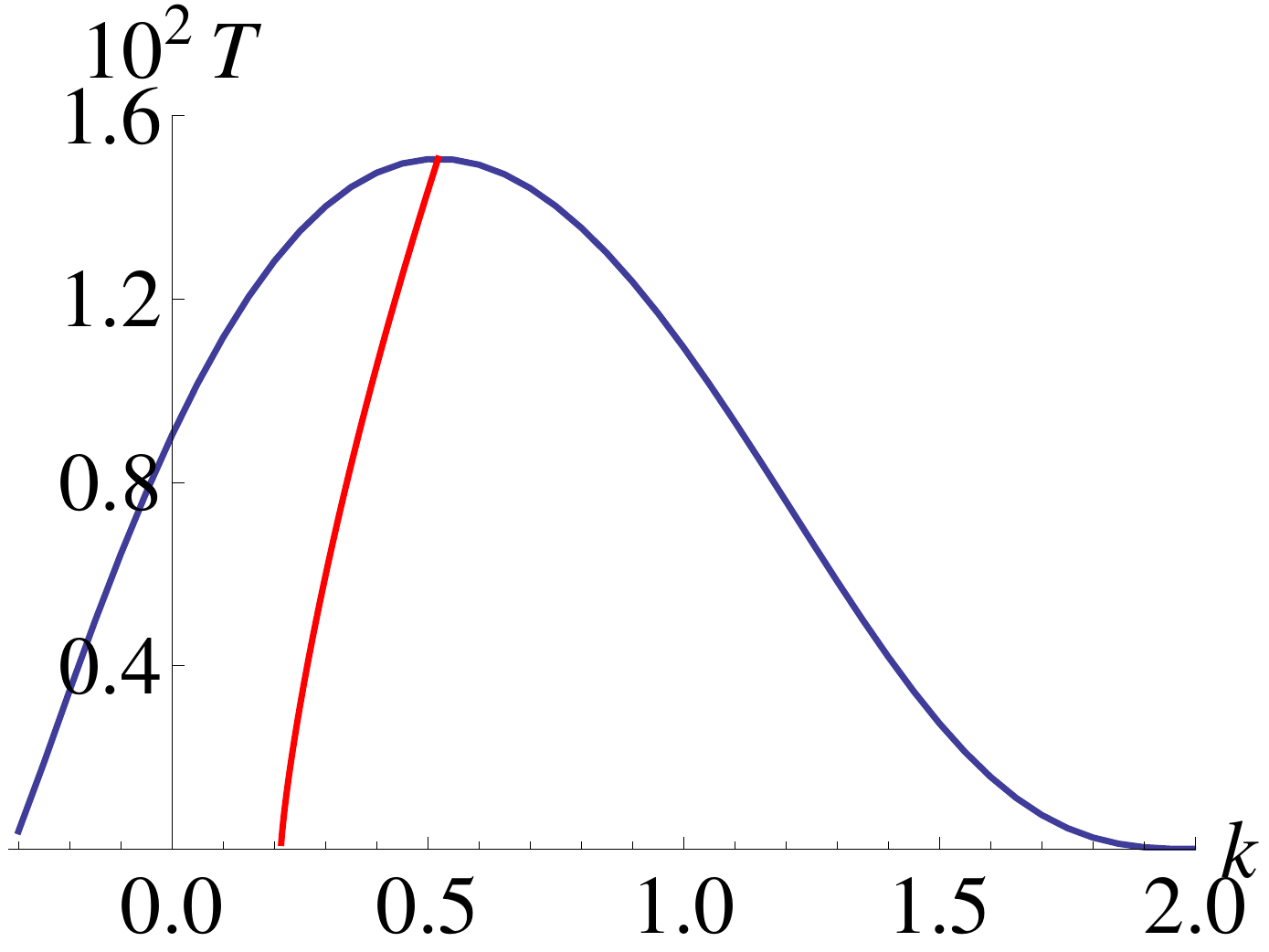}}\hskip 1 em
\subfloat[]{\includegraphics[width=4.5cm]{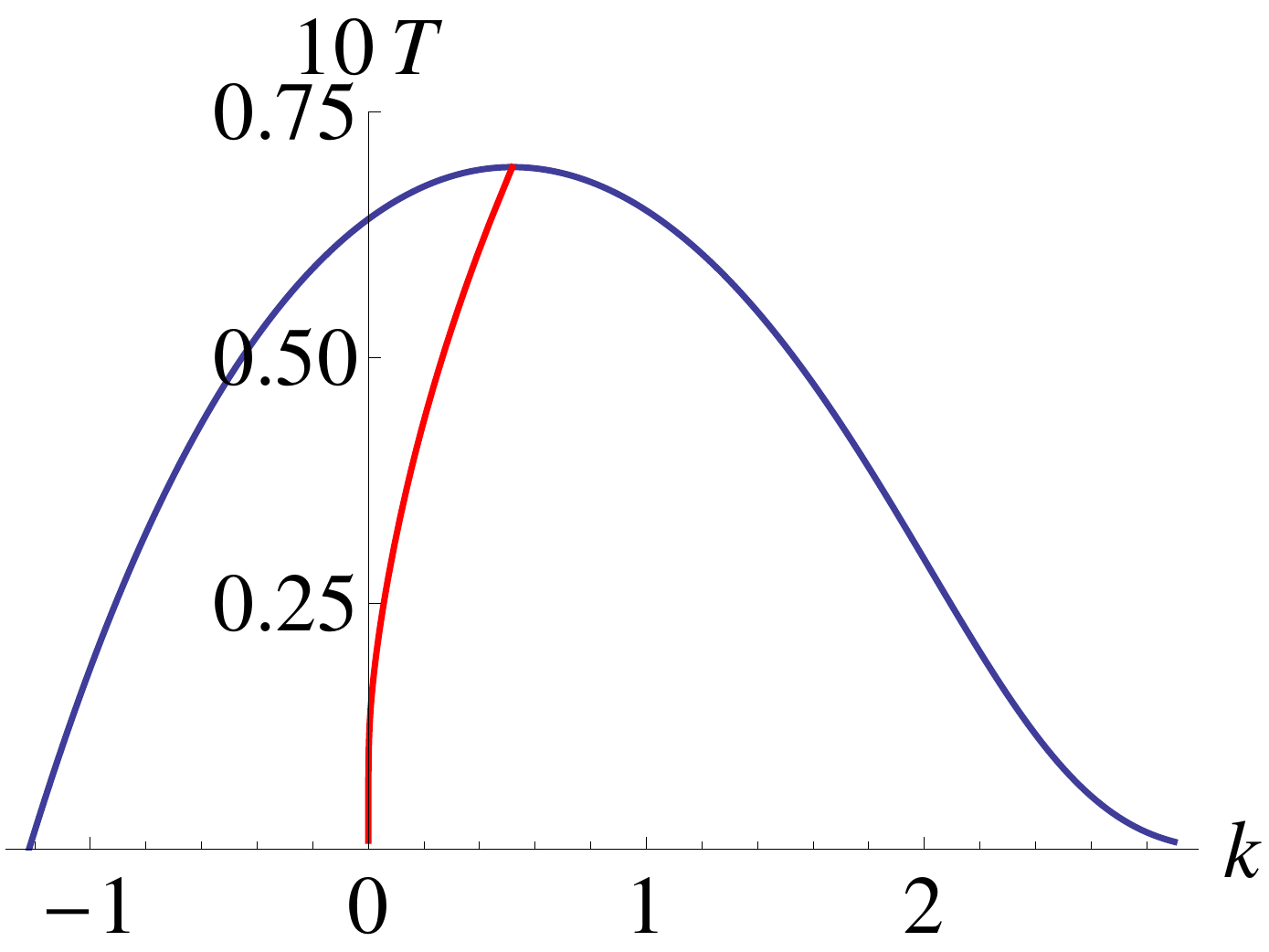}}\hskip 1 em
\subfloat[]{\includegraphics[width=4.5cm]{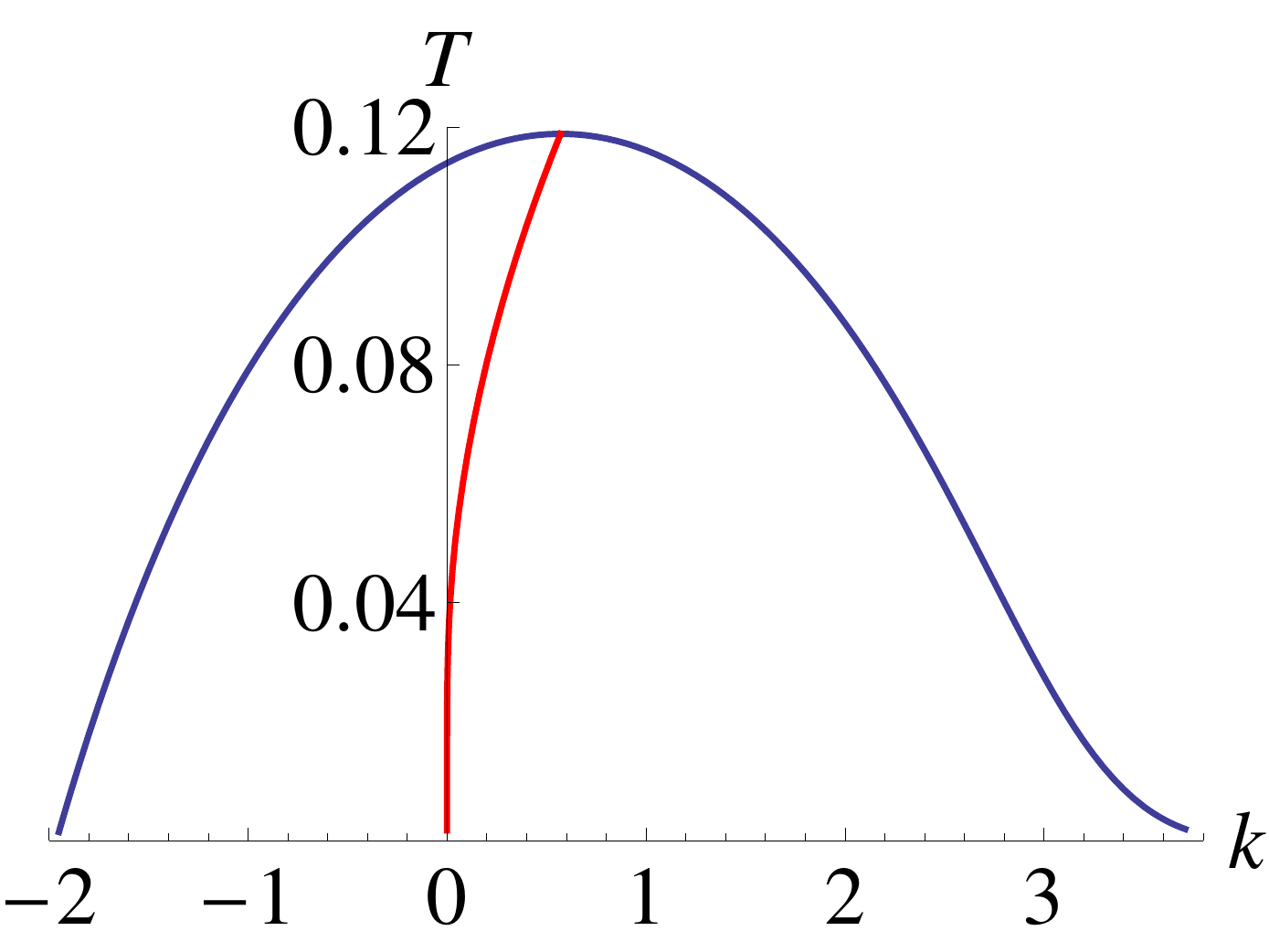}}
\caption{$(p+ip)$-wave black holes for $m=2$ and $e=2$ (panel (a)), $e=2.8$ (panel (b)) and $e=3.5$ (panel (c)). 
Each point under the bell curve corresponds to a $(p+ip)$-wave black hole
at temperature $T$ and wave-number $k$. All of these black holes have smaller free energy than the AdS-RN black hole at the same temperature.
The red curve in each figure corresponds to the thermodynamically preferred branch of black holes
that minimise the free energy density with respect to $k$ at fixed $T$.
For $e=3.5$ the $k=0$ axis is crossed by the red line at $T\sim 0.015$.
We have set $\mu=1$.}\label{figten}
\end{figure}

\begin{figure}
\centering
\subfloat[]{\includegraphics[width=5cm]{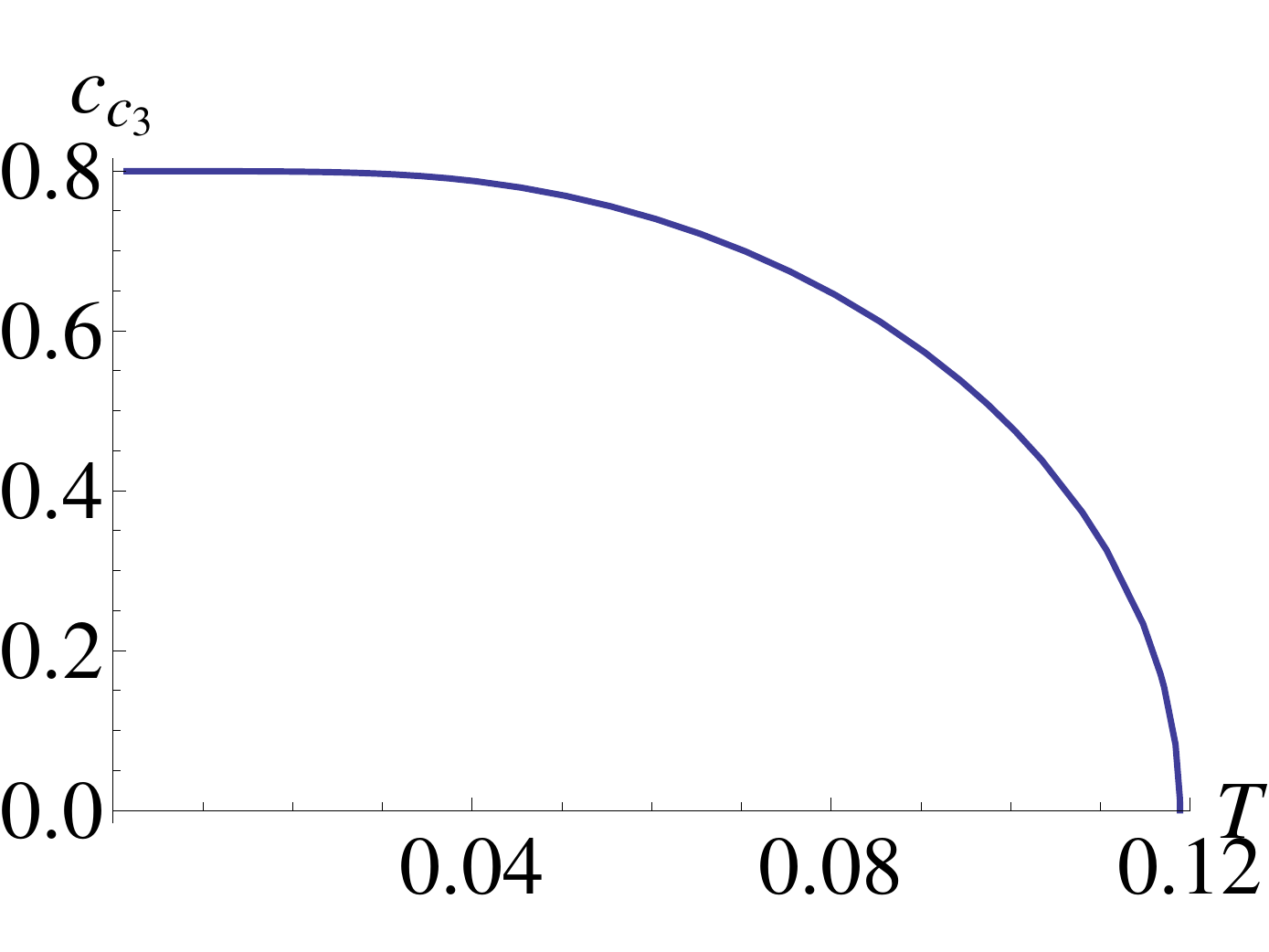}}\hskip 2 em
\subfloat[]{\includegraphics[width=5cm]{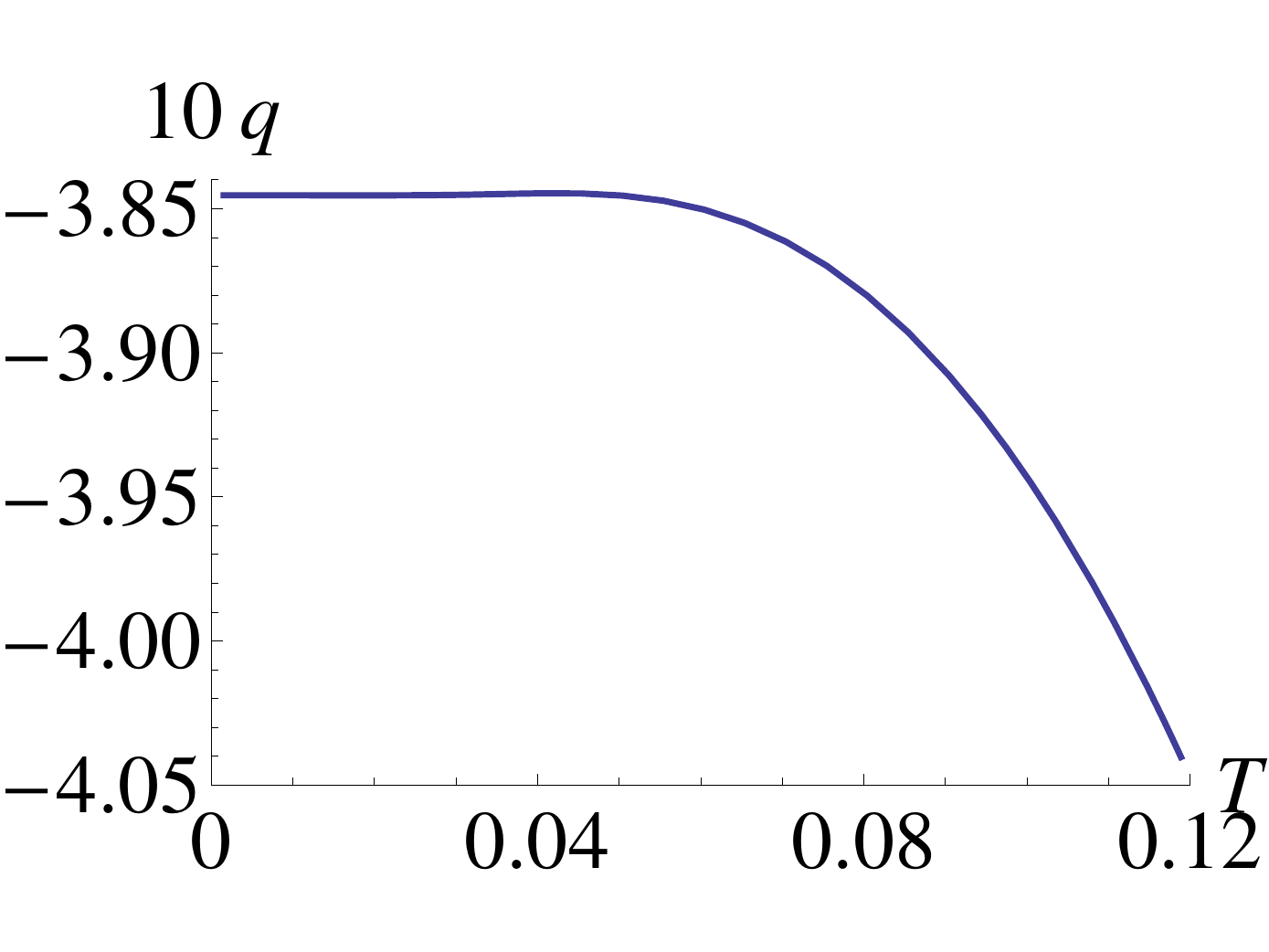}}
\caption{Properties of $(p+ip)$-wave black holes for $m=2$ and $e=3.5$ as a function of $T$ for the
thermodynamically preferred branch (the red line in figure \ref{figtwo}(c)).
Panel (a) plots $c_{c3}$ which fixes the $(p+ip)$-wave order parameter and panel (b)
plots $q$ which fixes the charge density. Note that these black holes have $c_b=c_Q=c_h=0$ and
the entropy density is displayed in figure \ref{figtwelve}.
We have set $\mu=1$.}\label{figeleven}
\end{figure}

We have numerically constructed these black holes for $m=2$ and various values of $e$, and we have summarised some of the results in figures \ref{figten} and \ref{figeleven}.
Some properties of these black holes, including their thermodynamics, will be discussed in the
following subsections. In figure \ref{figten} we display the two-parameter family of $(p+ip)$-wave black holes, corresponding to
the bell curves in figure \ref{figone}, including the thermodynamically preferred branches obtained by minimising the free-energy density
with respect to $k$ at fixed $T$, as we discuss below.  
Note that all black holes have smaller free energy than the AdS-RN black hole and
that the transition to the $(p+ip)$-wave preferred branch is second order.
In figure \ref{figeleven} we have plotted various physical quantities for the preferred branch
for the representative case of
$m=2$, $e=3.5$; other values of $e$ are similar. We will discuss the behaviour of the solutions as $T\to 0$ in section 
\ref{tzeropip}
below, where we will see that the black holes have zero entropy ground states.

It appears from the figures that the thermodynamically preferred black holes do not cross the $k=0$ axis, except possibly at $T=0$.
However, for the case $e=3.5$ the detailed numerics show that the $k=0$ axis is crossed at
temperatures of the order $T\sim 0.015$ and we expect this phenomenon to persist for larger values of $e$.

\subsection{Thermodynamics}
We analytically continue by setting $t=-i\tau$ . Near $r=r_+$, the Euclidean solution takes the approximate form
\begin{align}
\label{eq:ansatz2}
&ds_E^{2}\approx g_+ f_+^2(r-r_+)d\tau ^2+\frac{dr^2}{g_+(r-r_+)}+h_+^2(dx_1+\bar{Q}_+(r-r_+) d\tau)^2+r_+^2(dx_2^2+dx_3^3)\,,\nonumber\\
&A\approx \bar{a}_+(r-r_+) d\tau+b_+ dx_1\,,\nonumber\\
&C\approx e^{-i k x_1}(i\bar{c_1}d\tau+c_{2+}dr+ic_{3+} dx_1) \wedge (dx_2-i dx_3)\,,
\end{align}
where we have defined $Q_+=i \bar{Q}_+, a_+=i \bar{a}_+$ and $c_{1+}=i \bar{c_1}_+$ so that the metric, gauge-field and two-form are real. Regularity of the solution at $r=r_+$ is easily seen by making the coordinate change $\rho=2 g_+^{-1/2} (r-r_+)^{1/2} $ and making $\tau$ periodic with period $\Delta \tau=4 \pi /(g_+f_+)$ corresponding to temperature $T=(f_0 \Delta \tau)^{-1}$. We can also read off the area of the event horizon  and since we are working in units  with $ 16 \pi G=1$, we deduce that entropy density is given by 
\begin{equation}
s=4 \pi r^2 h_+\,.
\end{equation}
The total Euclidean action, $I_{Tot}=I+I_{ct}$ where $I=-iS$ and the counter-term action $I_{ct}$ is given in \eqref{ctermp}. In fact
the expression \eqref{ctermpex} for the $p$-wave ansatz is also valid for the $(p+ip)$-wave ansatz.
There are two equivalent ways to write the bulk part of the Euclidean action on-shell:
\begin{align}
I_{OS}=& Vol_3 \Delta\tau \int_{r_+}^\infty \left(2 r g h f\right)'\,,\nonumber \\
          =& Vol_3 \Delta\tau \int_{r_+}^\infty \bigg(r^2 h f g'+2 r^2 h g f'+r^2 \frac{h a}{f}(Q b'-a')-\frac{r^2 h^3}{f} QQ'\nonumber \\
&\qquad\qquad + c_3^2\frac{-ek a+m^2  h^2 Q+e^2 a b}{m[m^2 h^2 +(eb-k)^2]}-c_3 c_3'\frac{fgh}{[m^2 h^2 +(eb-k)^2]}\Bigg)'\,.
\end{align}
Notice that the first expression only receives contributions from the boundary at $r \to \infty$ since $g(r_+)=Q(r_+)=a(r_+)=0$, while the second expression also receives contributions from $r=r_+$. The free energy is defined by $W=T[I_{Tot}]_{OS}\equiv w Vol_3$.
Using the UV and the IR expansions \eqref{eq:UVexp}, \eqref{eq:IRexp}
we obtain the following expression for the free-energy density
\begin{align}
\label{eq:OSaction}
w&= -M\, ,
\end{align}
as well as the Smarr formula:
\begin{align}\label{smarrpipone}
4M +8 c_h+2 \mu q -s T=0\,.
\end{align}
Interestingly there are two further Smarr-type formulae that all of the $(p+ip)$-wave black holes satisfy, given by
\begin{align}\label{smarrtwo}
4 c_h-\frac{k}{e} c_b &=0\,,\nonumber\\
2c_Q+c_b \mu&=0\,.
\end{align}
A direct derivation is presented in appendix A. We can also obtain them using the results of 
\cite{Donos:2013cka}, as we discuss below.
 
A variation of the total on-shell action $[I_{Tot}]_{OS}$, for fixed $k$, gives
\begin{equation}
[\delta I_{Tot}]_{OS}=Vol_3 \Delta \tau [\delta f_0(3M +8 c_h +2 \mu q )+2 f_0 q \delta \mu]\,.
\end{equation}
 In this variation we are holding $\Delta \tau$ fixed and hence $\Delta \tau \delta f_0=-T^{-2} \delta T$. We next define the free energy $W$, and a corresponding density $w$, for the grand canonical ensemble via $W=T [I_{Tot}]_{OS}=w Vol_3$. We deduce that $w=w(T, \mu)$ with the first law given
by
\begin{equation}\label{flawpip}
\delta w=-s \delta T+2 q \delta \mu\,.
\end{equation}

We now compute the expectation value of the boundary stress-energy tensor. The relevant terms are again given by \eqref{stressy}
and using the asymptotic expansion \eqref{eq:UVexp}, we obtain
\begin{align}
\label{eq:bdySTpip}
\langle{T_{tt}}\rangle&=3 M +8 c_h\,,\nonumber\\
 \langle{T_{t x_1}}\rangle&=4 c_Q\,,\nonumber\\
\langle{T_{x_1 x_1}}\rangle&=M+8 c_h\,,\nonumber\\
\langle{T_{x_2 x_2}}\rangle&=M\,,\nonumber\\
\langle{T_{x_3 x_3}}\rangle&=M\,,
\end{align}
where we have set $f_0=1$.
One can easily check that this is traceless with respect to the flat boundary metric. 
We also notice that unlike the $p$-wave black holes, there is no spatial modulation of the stress tensor while on the other hand there
is, in general, momentum density in the $x_1$ direction. Defining the energy density $\varepsilon=3M +8 c_h$, we can rewrite $w=\varepsilon-T s+2 \mu q$
and thus the first law \eqref{flawpip} can be written $\delta \varepsilon=T \delta s -2 \mu \delta q$.

The relevant terms for calculating the expectation value of the current are as in \eqref{jterms}
and using \eqref{eq:UVexp}, we obtain
\begin{align}\label{jpip}
&\langle{J_{t}}\rangle=2q\,,\nn
&\langle{J_{x_1}}\rangle=2 c_b\,,
\end{align}
where we set $f_0=1$. From the temporal component we see that the constant $q$ fixes the charge density and that, in general,
there is current density in the $x_1$ direction.

\subsubsection{Variation of $w$ with respect to $k$}
For a given temperature we are interested in the solution labelled by $k_{min}$ which minimises
the free energy. These black holes are specified by varying the action $I_{Tot}$ with respect to $k$, putting it on-shell and then 
setting it to zero. We first note that
\begin{equation}
k\,\partial_{k}I_{Tot}=-Vol_3\Delta\tau\int^\infty_{r_+} dr\,  \frac{2k}{m}c_1 c_2\,.
\end{equation}
One of the components of the gauge-field equation of motion, given in
\eqref{from}, implies that this integrand is a total derivative with respect to $r$,
and hence can be easily evaluated. Substituting the asymptotic and near horizon expansions 
\eqref{eq:UVexp},\eqref{eq:IRexp}
we find that at constant $T$ we have 
\begin{align}\label{extp}
k\partial_k w=\frac{2k}{e}c_b\,.
\end{align}
We thus deduce that
\begin{align}
\partial_k w=0\quad\Rightarrow\quad c_b=c_h=c_Q=0\,.
\end{align}
where we used the Smarr formulae \eqref{smarrtwo}.
In particular we see from \eqref{eq:bdySTpip} that the dual stress tensor for this one-parameter family of black hole solutions 
is homogeneous and isotropic with no momentum density in the $x_1$ direction. In addition, from \eqref{jpip} we see
that the current density in the $x_1$ direction vanishes.

\subsubsection{Connection with \cite{Donos:2013cka}}
The general results of \cite{Donos:2013cka} for the thermodynamics of periodic black holes imply, in the present set-up, that
\begin{align}\label{eq:final_var_illpip}
w&=-Ts-\bar J^{t}\mu+\bar T^{tt}\,,\nn
w&=-\bar T^{x_2x_2}=-\bar T^{x_3x_3}\,,\nn
w&=-T^{x_1x_1}-J^{x_1}\bar a_{x_1}\,,\nn
0&=-T^{x_1 t}+\mu \bar J^{x_1}\,,\nn
\delta w&=-\bar J^{t}\delta \mu-s{\delta T}+\frac{\delta k}{k}\left(w+T^{x_1x_1}\right)\,,
\end{align}
where $a_{x_1}$ is a source term for the current and 
the bars refer to quantities averaged over a period in the $x_1$ direction, as in section \ref{conp}, and we have used the fact
that stress energy conservation and current conservation imply $T^{x_1x_1}$, $T^{tx_1}$ and $J^{x_1}$ are constants.
Using the actual expressions for the stress tensor and current given in \eqref{eq:bdySTpip}, \eqref{jpip} and substituting into 
\eqref{eq:final_var_illpip} we find that
\begin{align}\label{eq:final_var_illsec}
w&=-Ts+2q\mu+3M+8c_h\,,\nn
w&=-M\,,\nn
w&=-M-8c_h+\frac{k}{e}2c_b\,,\nn
0&=4c_Q+2 c_b\mu\,,\nn
\delta w&=2q\delta \mu-s{\delta T}+\frac{\delta k}{k}\left(w+M+8c_h\right)\,,
\end{align}
thus recovering various expressions that we derived earlier.
It is worth noting that in obtaining the third expression, which when combined with the second gives one of the Smarr-type formula in \eqref{smarrtwo},
we worked in the gauge where there was no phase in the two-form $C$ but a source term appearing in the gauge-field: i.e. $a_{x_1}=-k/e$. 
If we wanted to work in the original gauge, with $a_{x_1}=0$, we would need to extend the formalism of \cite{Donos:2013cka} to suitably take into account
gauge-invariant variables.

\subsubsection{Solutions with $k=0$}
We briefly comment on the $k=0$ branch of black holes. 
Recall that for suitably large $e$, such as $e=3.5$, the thermodynamically preferred branch seems to cross the $k=0$ 
axis at very low temperatures. 
When $k=0$, we deduce from \eqref{smarrtwo} that $c_h=0$. On the other hand, we find from our numerical results
that $c_Q$ and $c_b$ are non-zero. 
In particular, from \eqref{extp} we see that $c_b=0$ for the $k=0$
branch show that they are unstable.
Unlike the $p$-wave case, there is no enhanced symmetry for the $k=0$ black holes
and there is no additional consistent truncation.

\subsection{Behaviour as $T\to 0$}\label{tzeropip}
\begin{figure}
\centering
\subfloat[]{\includegraphics[width=4.5cm]{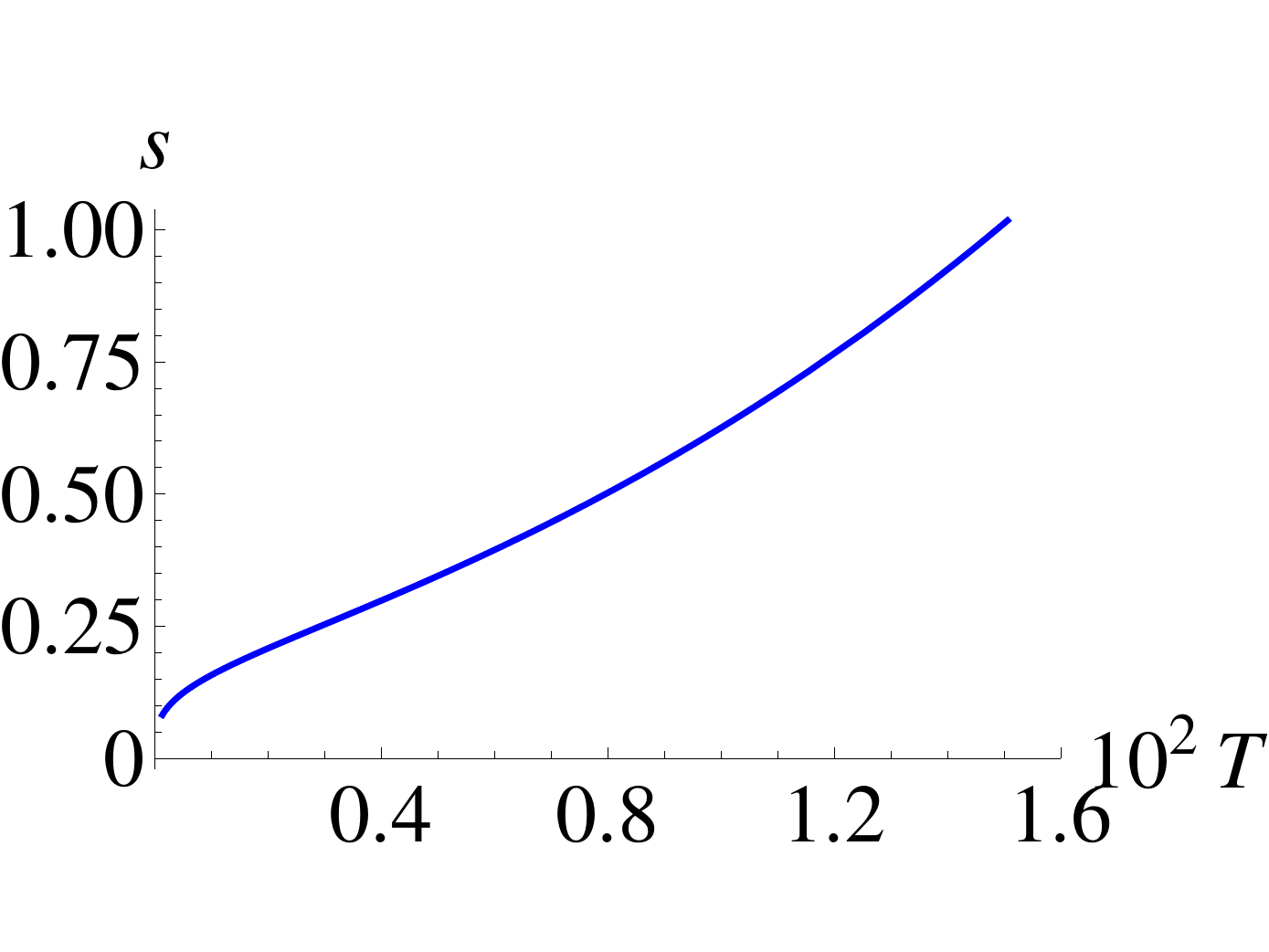}}\hskip1em
\subfloat[]{\includegraphics[width=4.5cm]{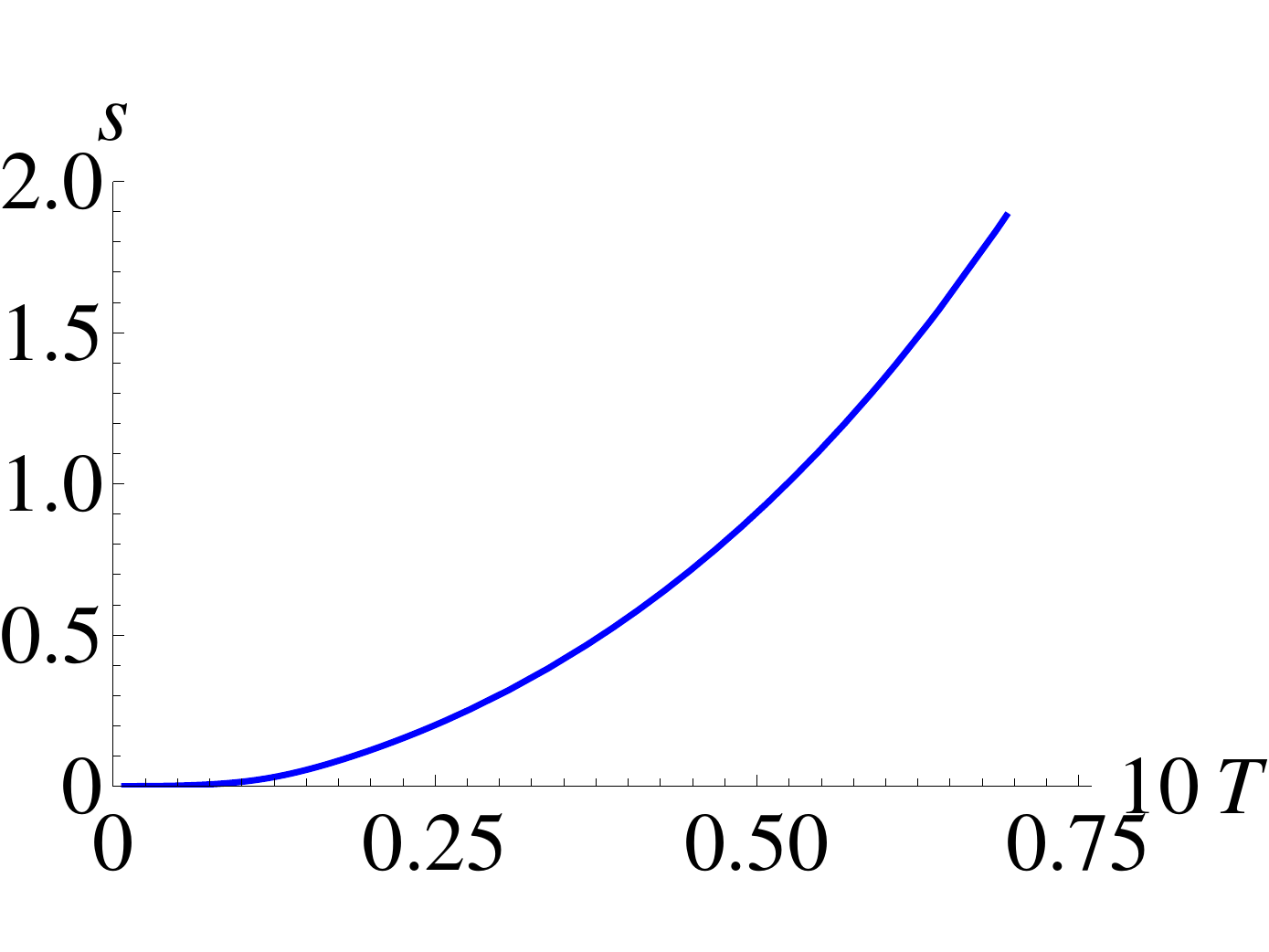}}\hskip1em
\subfloat[]{\includegraphics[width=4.5cm]{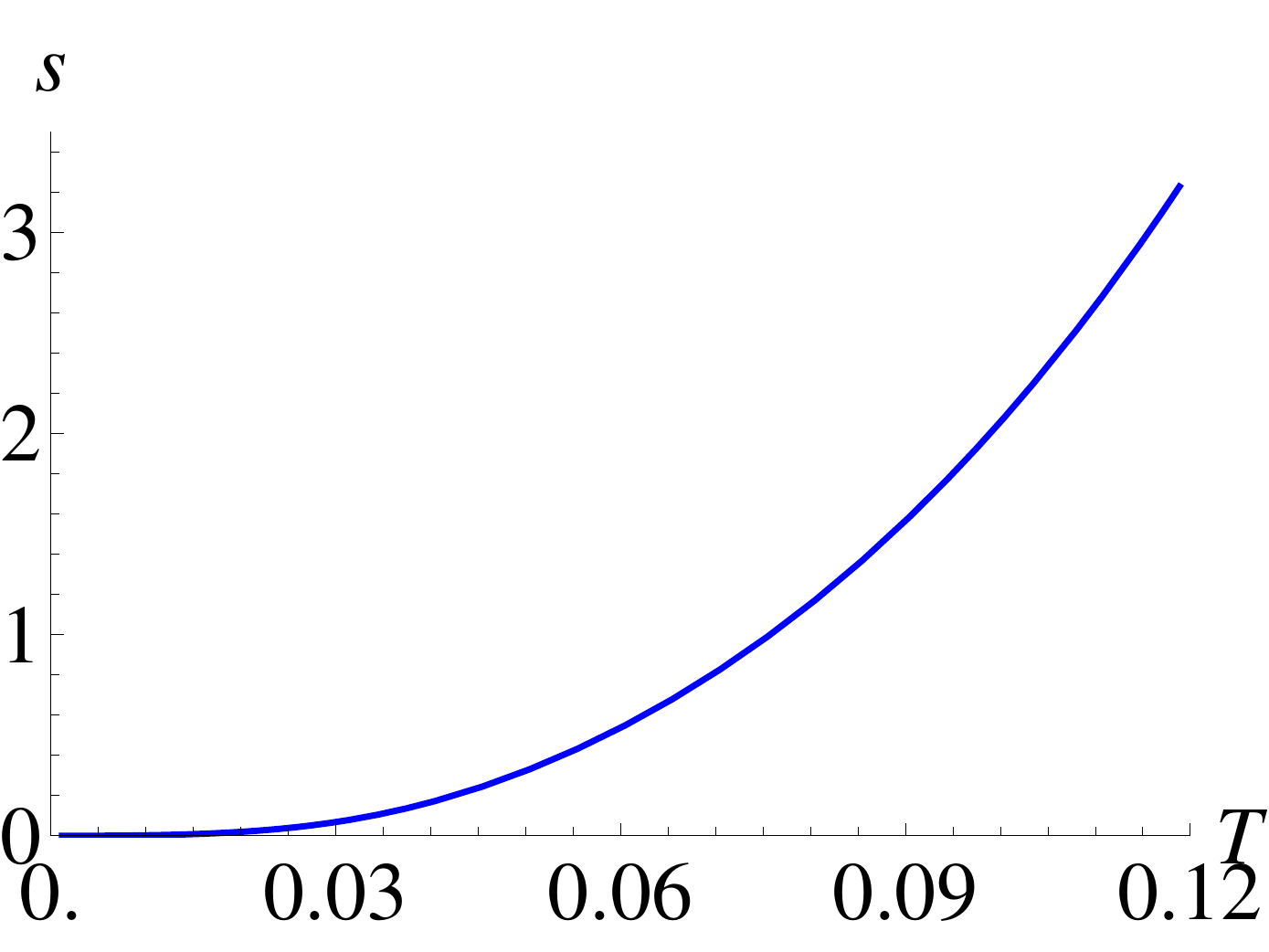}}
\caption{The low temperature behaviour of the entropy density $s$ for the 
$(p+ip)$-wave black holes. The plots are for the thermodynamically preferred black holes in figure \ref{figtwo}
with $m=2$ and $e=2$ (figure a), $e=2.8$ (figure b) and $e=3.5$ (figure c). Observe that for $e=2$ the zero entropy ground state
behaviour only starts to appear for temperatures of the order $10^{-3}$. 
We have set $\mu=1$.}\label{figtwelve}
\end{figure}

For all of the $(p+ip)$-wave black holes we find that $s\to 0$ as $T\to 0$, as we see in figure \ref{figtwelve}.
It is worth noting that in some cases, for example $m=2$, $e=2$, one has to go to temperatures of the order $10^{-3}$ to see 
that the entropy density is in fact going to zero.

As yet, we have not been able to elucidate the nature of the ground state solutions as $T\to 0$. In particular, there do not seem
to be analogues of the scaling solutions that we saw in the $p$-wave case, that we discussed in sections \ref{scalhelp} and \ref{scalkzp}. However, it seems to be possible to construct domain wall solutions interpolating between $AdS_5$ and itself
using a similar expansion to what we considered in section \ref{pads5ir}, and these may play a key role.
Indeed as $r\to 0$, we consider
\begin{align}
&g=r^2+\delta g,\qquad f=\bar f_0+\delta f,\qquad h=h_0 r+\delta h,\nn
&Q=Q_0+\delta Q,\qquad c_3=\delta c_3,\qquad a=a_0+\delta a, \qquad b=b_0+\delta b_0\,,
\end{align}
where $\bar f_0, h_0,Q_0,a_0$ and $b_0$ are constants. After substituting into the equations of motion we find that at first order in
the perturbation we just need to solve a second order linear ODE for $\delta c_3$. We find that
the following solution can be developed:
\begin{align}
\delta c_3=c_3^0\frac{e^{-{kx}/h_0r}}{r^{1/2}}(1+{\cal O}(r))\,,
\end{align}
where 
\begin{align}
x=\frac{(k-w_0 e)}{k}\sqrt{1-\frac{h_0^2\left(a_0 e+Q_0(k-b_0 e)\right)^2}{f_0^2(k-b_0 e)^2}}\,.
\end{align}
At next order in the perturbation we will get terms with $e^{-{2kx}/h_0r}$ appearing. This expansion is specified by six IR parameters.
Combined with the seven UV parameters and two scaling symmetries from \eqref{eq:symmetries}, 
we expect to find domain wall solutions that depend on
one-parameter which can be taken to be $k$.

Finally, we point out that we can use the argument presented in section \ref{littlearg} to conclude that $\partial_T k=0$ as $T\to 0$. 
Indeed this is the behaviour we see in figures \ref{figten}(b) and \ref{figten}(c).
This is not so immediate for figure \ref{figten}(a) but further evidence can be obtained using a more detailed analysis of the low temperature limit.

\section{$p$-wave versus $(p+ip)$-wave order}
We have constructed $p$-wave and $(p+ip)$-wave black holes for $m=2$ and various values of $e$.
In both cases the black holes have smaller free energy than that of the AdS-RN black holes and the transition
from the AdS-RN phase to the thermodynamically preferred phase is second order, in each case.

\begin{figure}
\centering
\subfloat[]{\includegraphics[width=8cm]{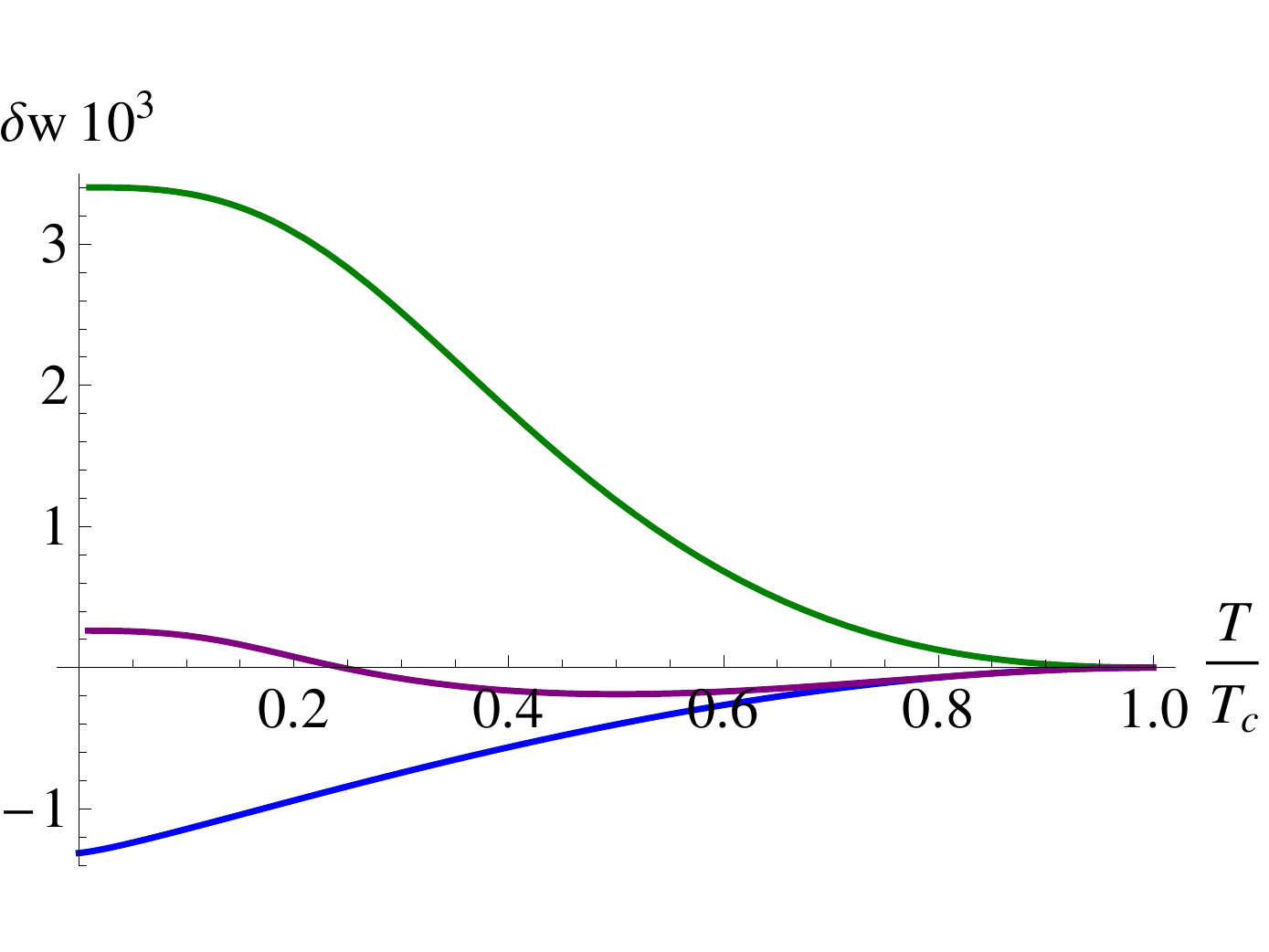}}\hskip1em
\caption{A plot of the difference, $\delta w$, of the free energy density of the $p$-wave and the $(p+ip)$-wave branches of thermodynamically preferred 
black holes (the red lines in figures \ref{figtwo} and \ref{figten}). 
The plots are for $m=2$. The blue curve is for $e=2$ and shows that the $p$-wave order is preferred for all $T\le T_c$.
The purple curve is for $e=2.8$ and shows that the $p$-wave order is first preferred and then there is
a first order transition at $T\sim 0.245 T_c$ to the $(p+ip)$-wave order. The green curve is for $e=3.5$ and shows the $(p+ip)$-wave order
is preferred for all $T\le T_c$.
We have set $\mu=1$.}\label{figthirteen}
\end{figure}

We now compare the free energies for the $p$-wave and the $(p+ip)$-wave black holes.
For $e=2$ we see in figure \ref{figthirteen} that the $p$-wave black holes
are preferred all the way down to zero temperature. As the value of $e$ is increased, 
the $(p+ip)$-wave become more favourable.
Our numerical results indicate that at some critical value of $e$, whose precise value is hard to establish, the branches meet
at $T=0$ (i.e. they have the same free energy - the solutions seem to be distinct). Increasing $e$ a little further we
get to the situation exemplified by $e=2.8$ in figure \ref{figthirteen}. In this case there is a second order phase transition from the 
AdS-RN black holes to 
the $p$-wave black holes, followed by a first order transition to the $(p+ip)$-wave black holes. 
Beyond another critical value of $e$ the $p$-wave branch is never preferred and there is
just a second order transition from the AdS-RN black holes to the $(p+ip)$-wave black holes. This is illustrated for $e=3.5$ in figure \ref{figthirteen}. 

The phase structure that we have found for $m=2$ and various $e$ implies that the phenomenon of pitch inversion that we saw for the
$p$-wave branches is not thermodynamically preferred.  
Furthermore, for $m=2$ we find that of the three types of $T=0$ ground states for the $p$-wave black holes which we discussed in sections \ref{scalp} and \ref{tzerolimp},
only the Bianchi VII$_0$ scaling solutions are preferred. It is quite possible that the situation changes for different values of $m$.
Another open issue is whether the $p$-wave and $(p+ip)$-wave black holes themselves are unstable. If they are, then 
they will sprout additional black hole branches (as in \cite{Donos:2011ut}) which could modify the conclusions concerning the phase structure.

\section{Discussion}
Building on \cite{Donos:2011ff,Donos:2012gg} we have investigated in further detail the $p$-wave and $(p+ip)$-wave superconducting black holes
that arise in a $D=5$ theory of gravity coupled to a gauge-field and a charged two-form. The model depends on two-parameters, $m$ and
$e$ and we focussed on the case $m=2$ and then varied $e$. For the $p$-wave black holes, for some values of $e$ 
we again saw some of the features seen in \cite{Donos:2012gg} for $m=1.7$ and $e=1.88$. 
For example, when $e=2$ the zero temperature black holes approach 
ground states which are domain walls interpolating between $AdS_5$ in the UV and a Bianchi VII$_0$ scaling solution in the IR.
For other values of $e$ we saw some new features. Firstly, when $e\gtrsim 2.9$ the thermodynamically preferred 
black holes exhibit the phenomenon of pitch inversion. Secondly, when $e\sim 2.9$ the zero temperature limits of the black holes appear to approach domain walls interpolating from $AdS_5$ in the UV and a novel anisotropic scaling solution in the IR with $k=0$. 
For larger values of $e$ the ground state solutions appear to interpolate between $AdS_5$ in the UV and $AdS_5$ in the IR.
It will be particularly interesting to explore this latter class of domain walls in more detail. For example, based on
\cite{Horowitz:2009ij}, one might expect that they will have conductivities with novel properties.

The back-reacted $(p+ip)$-wave black holes, depending on wave number $k$ that we constructed here are new. 
The $(p+ip)$-wave black holes
also approach zero temperature ground states with vanishing entropy. While we have not been able to identify the ground states
that occur in this case, we argued that there are possible domain wall solutions interpolating between the same
$AdS_5$ in the UV and the IR, which may play an important role. We analysed the competition between the different orders
and found that for low values of $e$ the $p$-wave order is preferred, but for higher values $(p+ip)$-wave order is preferred. For 
intermediate values of $e$ there is a first order transition from the $p$-wave to the $(p+ip)$-wave order.

Finally, it would be interesting to carry out similar investigations into the $p$-wave and $(p+ip)$-wave black holes
that arise in theories of gravity coupled to $SU(2)$ gauge-fields extending
\cite{Gubser:2008zu,Gubser:2008wv,Roberts:2008ns,Ammon:2009xh}. In particular, we expect 
that analogously rich classes of black solutions and ground states will be present for the $D=5$ models studied in
\cite{Donos:2011ff} (that developed the work of \cite{Zayas:2011dw}). It would also be of interest to investigate the interplay
of helical magnetic orders and superconductivity, that have been discussed in \cite{JPSJ.80.114712,2013PhRvB..87l1112C}, within a holographic context.

\section*{Acknowledgements}
We thank A. Green, S. Hartnoll and J. Sonner for helpful discussions. 
The work is supported by STFC grant ST/J0003533/1
and also by the European Research Council under the European Union's Seventh Framework Programme (FP7/2007-2013), ERC Grant agreement STG 279943, ``Strongly Coupled Systems".
CP is supported by an I.K.Y. Scholarship. 

\appendix
\section{Smarr formulae for the $(p+ip)$-wave black holes}
For an arbitrary Killing vector $\zeta^a$ we have
\begin{equation}
R^a{}_{b}\zeta^b=\nabla_b\nabla^a \zeta^b\,.
\end{equation}
Since the metric ansatz \eqref{eq:ansatz} depends only on the radial coordinate $r$, this can be rewritten
\begin {equation}
R^a{}_b\zeta^b=-\frac{1}{\sqrt{-g}}\partial_r[\sqrt{-g} \nabla^{r}\zeta^{a}]\,,
\end{equation}
and we obtain
\begin{align}
\label{eq:LHS}
\sqrt{-g} R^t{}_t&=-\partial_r(r^2 ghf'+\frac{1}{2}r^2fhg'-\frac{r^2 h^3 Q Q'}{2f})\,,\nonumber\\
\sqrt{-g} R^{x_1}{}_t&=\partial_r(r^2 g h Qf'+\frac{1}{2}r^2 f h Q g'-r^2 fgQh'-\frac{1}{2} r^2 f g h Q'-\frac{r^2 h^3 Q^2 Q'}{2f})\,,\nn
\sqrt{-g} R^{x_1}{}_{x_1}&=-\partial_r(r^2 fgh'+\frac{r^2 h^3 Q Q'}{2f})\,,\nonumber\\
\sqrt{-g} R^{x_2}{}_{x_2}&=-\partial_r(r fgh)\,,
\end{align}
where the first two equations arise from the Killing vector $\partial_t$ and the next two equations arise
from the Killing vectors $\partial_{x_1}$ and $\partial_{x_2}$, respectively. As in \cite{Bhattacharya:2011eea} 
the strategy to obtain the
Smarr formula is to use the equations of motion \eqref{fulleom} to find linear combinations of these four equations
so that the left hand side is also a total derivative.

We first write the Einstein equations appearing in \eqref{fulleom} as
\begin{align}
R_{\mu\nu}=-4g_{\mu\nu}+X^{(F)}_{\mu\nu}+X^{(C)}_{\mu\nu}\,,
\end{align}
where
\begin{align}
X^{(F)}_{\mu\nu}&=\tfrac{1}{2}\left(F_\mu{}^\rho F_{\nu\rho}-\tfrac{1}{6}g_{\mu\nu}F_{\rho\sigma}F^{\rho\sigma}\right)\,,\nn
X^{(C)}_{\mu\nu}&=\tfrac{1}{2}\left(C_{(\mu}{}^\rho\bar C_{\nu)\rho}-\tfrac{1}{6}g_{\mu\nu}C_{\rho\sigma}\bar C^{\rho\sigma}\right)\,.
\end{align}
Let us focus on the components $X^{(F)i}{}_j$ where $i,j$ run over the coordinates $t,x_1,x_2,x_3$. Since
the gauge-field in our ansatz has the form $A=a_i(r)dx^i$ we have
\begin{align}
\sqrt{-g}X^{(F)i}{}_j&=\tfrac{1}{2}\sqrt{-g}\left(F^{ri}\delta^k_j-\tfrac{1}{3} F^{rk}\delta^i_j\right)\partial_r a_k\,,\nn
&=\partial_r \left[\tfrac{1}{2}\sqrt{-g}\left(F^{ri}\delta^k_j-\tfrac{1}{3} F^{rk}\delta^i_j\right) a_k\right]\nn
&\qquad\qquad
-\tfrac{1}{2}\left[\partial_r\left(\sqrt{-g}F^{ri}\right)\delta^k_j-\tfrac{1}{3} \partial_r\left(
\sqrt{-g}F^{rk}\right)\delta^i_j\right]a_k\,.
\end{align}
We can then rewrite the last line using the following equations of motion for the gauge-field
\begin{align}\label{terms}
\partial_r\left(\sqrt{-g}F^{ri}\right)=-\frac{e}{4m}\sqrt{-g}\epsilon^{irjkl}\left[C_{rj}\bar C_{kl}+\bar C_{rj}C_{kl}\right]\,.
\end{align}
The right hand side of this expression contains terms that are quadratic in the functions
$c_1, c_2$ and $c_3$, as does $\sqrt{-g}X^{(C)i}{}_j$. In fact the only non-trivial relations in \eqref{terms}  
are given by
\begin{align}\label{from}
\partial_r\left(\sqrt{-g}F^{rt}\right)=-\frac{2e}{m}c_2c_3\,,\nn
\partial_r\left(\sqrt{-g}F^{rx_1}\right)=\frac{2e}{m}c_1c_2\,.
\end{align}
The last step is to use the equation of motion for
the two-form $C$ to find linear combinations where the terms that are quadratic in the $c_i$ can be expressed
as a total derivative with respect to $r$. 
The $(r,t,x_2)$ and $(r,x_1,x_2)$ components of the equation of motion $H=-im*C$ can be written as
\begin{align}
\partial_r c_1&=eac_2-\frac{m}{fgh}\left(h^2Qc_1+(f^2g-h^2 Q^2)c_3\right)\,,\nn
\partial_r c_3&=(eb-k)c_2-\frac{mh}{fg}\left(c_1-Qc_3\right)\,.
\end{align}
It is also useful to note that the equation of motion for $C$ also implies that $c_2$ can be solved
algebraically in terms of $c_1$ and $c_3$:
\begin{align}
c_2=\frac{1}{mfgh}\left(eac_3-(eb-k)c_1\right)\,,
\end{align}
as one can deduce from \eqref{eq:c1c2}.

After combining these ingredients, it is straightforward to show that the relevant linear combinations are given by
\begin{align}
\label{eq:RHS}
\sqrt{-g} (R^t{}_t-R^{x_2}{}_{x_2})&=\partial_r\left(\frac{1}{2}\sqrt{-g} a F^{rt}+\frac{1}{2m} c_1c_3\right)\,,\nn
\sqrt{-g} (R^{x_1}{}_{x_1}-R^{x_2}{}_{x_2})&=\partial_r\left(\frac{1}{2}\sqrt{-g} (b-\frac{k}{e})F^{rx_1 }-\frac{1}{2m} c_1c_3\right)\,,\nn
\sqrt{-g} R^{x_1}{}_t&=\partial_r\left(\frac{1}{2}\sqrt{-g} aF^{rx_1 }-\frac{1}{2m} c_1c_1\right)\,.
\end{align}
Combining equations \eqref{eq:LHS} with \eqref{eq:RHS} we can then integrate from $r=r_+$ to
$r=\infty$. After substituting in the asymptotic \eqref{eq:UVexp}
and near horizon \eqref{eq:IRexp} expansions, we obtain three Smarr-type formulae
given by
\begin{align}
\label{eq:result}
4M +8 c_h+2 \mu q -s T&=0\,,\nonumber\\
4 c_h-\frac{k}{e} c_b &=0\,,\nonumber\\
2c_Q+c_b \mu&=0\,.
\end{align}
where we have set $f_0=1$. These are the formulae \eqref{smarrpipone} and \eqref{smarrtwo} stated in the main text for the $(p+ip)$-wave black holes.
\bibliographystyle{utphys}
\bibliography{helical}{}
\end{document}